\documentclass[prd,twocolumn,floatfix,preprintnumbers,showpacs]{revtex4}
\usepackage{graphicx}
\usepackage{dcolumn}
\usepackage{bm}
\usepackage{color}
\usepackage{hyperref}

\def\lsim{\mathrel{\mathop
  {\hbox{\lower0.5ex\hbox{$\sim$}\kern-0.8em\lower-0.7ex\hbox{$<$}}}}}
\def\gsim{\mathrel{\mathop
  {\hbox{\lower0.5ex\hbox{$\sim$}\kern-0.8em\lower-0.7ex\hbox{$>$}}}}}

\begin{document}
\newcommand{\mincir}{\raise
-2.truept\hbox{\rlap{\hbox{$\sim$}}\raise5.truept 
\hbox{$<$}\ }}
\newcommand{\magcir}{\raise
-2.truept\hbox{\rlap{\hbox{$\sim$}}\raise5.truept
\hbox{$>$}\ }}
\newcommand{\minmag}{\raise-2.truept\hbox{\rlap{\hbox{$<$}}\raise
6.truept\hbox
{$>$}\ }}

\newcommand{\Ocdm}{\Omega_{\rm cdm}}
\newcommand{\ocdm}{\omega_{\rm cdm}}

\input epsf

\preprint{LAPTH-1179/07, astro-ph/0703625} 
\title{New constraints on the observable inflaton potential 
from WMAP and SDSS} 
\author{Julien
Lesgourgues}\email{julien.lesgourgues@lapp.in2p3.fr}\author{Wessel
Valkenburg}\email{wessel.valkenburg@lapp.in2p3.fr} \affiliation{ LAPTH\footnote{Laboratoire de Physique
Th\'eorique d'Annecy-le-Vieux, UMR5108}, Universit\'e de Savoie \&
CNRS, 9 chemin de Bellevue, BP110, F-74941 Annecy-le-Vieux Cedex,
France } \date{\today}
\pacs{98.80.Cq}
\begin{abstract}
We derive some new constraints on single-field inflation from the
Wilkinson Microwave Anisotropy Probe 3-year data combined with the
Sloan Luminous Red Galaxy survey. Our work differs from previous
analyses by focusing only on the observable part of the inflaton
potential, or in other words, by making absolutely no assumption about
extrapolation of the potential from its observable region to its
minimum (i.e., about the branch of the potential responsible for the
last $\sim$50 inflationary e-folds). We only assume that inflation
starts at least a few e-folds before the observable Universe leaves
the Hubble radius, and that the inflaton rolls down a monotonic and
regular potential, with no sharp features or phase transitions. We
Taylor-expand the inflaton potential at order $v=2$, 3 or 4 in the
vicinity of the pivot scale, compute the primordial spectra of scalar
and tensor perturbations numerically and fit the data. For $v > 2$, a
large fraction of the allowed models is found to produce a large
negative running of the scalar tilt, and to fall in a region of
parameter space where the second-order slow-roll formalism is strongly
inaccurate. We release a code for the computation of inflationary
perturbations which is compatible with {\sc cosmomc}.
\end{abstract}

\maketitle

Cosmological inflation is known to be a successful paradigm providing
self-consistent initial conditions to the standard cosmological
scenario
\cite{Starobinsky:1980te,Guth:1980zm,Sato:1980yn,Hawking:1981fz,Linde:1981mu,Linde:1983gd}
and explaining the generation of primordial cosmological perturbations
\cite{Starobinsky:1979ty,Hawking:1982cz,Starobinsky:1982ee,Guth:1982ec,Linde:1982uu,Bardeen:1983qw,Abbott:1984fp,Salopek:1988qh}.
The distribution of Cosmic Microwave Background (CMB) anisotropies, as
observed for instance by the Wilkinson Microwave Anisotropy Probe (WMAP)
\cite{Spergel:2006hy,Page:2006hz,Hinshaw:2006ia,Jarosik:2006ib}, is
compatible with the simplest class of inflationary models called {\it
single-field inflation.}


The definition of single-field inflation is not unique: for instance,
some authors consider hybrid inflation
\cite{Linde:1991km,Linde:1993cn,Copeland:1994vg} as a multi-field
model, since it involves one scalar field in addition to the inflaton field
(the role of the second field being to trigger the end of
inflation). In this work, we call {\it single-field inflation} any
model in which the observable primordial spectrum of scalar and tensor
metric perturbations can be computed using the equation of motion of a
single field. This definition does include usual models of hybrid
inflation.


The goal of this paper is to derive from up-to-date cosmological data
some constraints on the scalar potential $V(\phi)$ of
single-field inflation. This question has already been addressed in
many interesting works since the publication of WMAP 3-year results
\cite{Spergel:2006hy,Peiris:2006ug,deVega:2006hb,Easther:2006tv,Kinney:2006qm,Martin:2006rs,Covi:2006ci,Finelli:2006fi,Peiris:2006sj,Destri:2007pv,Ringeval:2007am,Cardoso:2006wf} (see also \cite{Cline:2006db} for earlier results).
Our approach is however different, since all these references assume
either that the slow-roll formalism can be applied (at first or second
order), or that the scalar potential can be extrapolated from the
region directly constrained by the data till the end of inflation. We
want to relax these two restrictions simultaneously, and to derive
constraints on the {\it observable part of the inflaton potential}
under the only assumption that $V(\phi)$ is smooth enough for being
Taylor-expanded at some low order in the region of interest. In this
respect, our work is still not completely general and does not explore
possible sharp features in the inflaton potential (see
e.g. \cite{Martin:2006rs,Covi:2006ci} for recent proposals).
Throughout the abundant literature on the inflaton potential
reconstruction, the work following the closest methodology to ours is
the pre-WMAP paper of Grivell and Liddle \cite{Grivell:1999wc}.

%
%

The question of whether the slow-roll formalism can be safely employed
or not is intimately related to the magnitude of a possible {\it
running of the tilt} in the primordial spectrum of curvature
perturbations. In order to clarify this point, lets us first recall
that the slow-roll
formalism~\cite{Steinhardt:1984jj,Salopek:1990jq,Liddle:1994dx}
consists in employing analytical expressions for the primordial
spectrum of curvature perturbations ${\cal P}_{\cal R}(k)$ and
gravitational waves ${\cal P}_{h}(k)$. Such expressions hold in the
limit in which the first and second logarithmic derivative of the
Hubble parameter $H$ with respect to the e-fold number $N \equiv \ln
a$ remain smaller than one throughout the $\Delta N \sim 10$
observable e-folds of inflation (i.e., over the period during which
observable Fourier modes cross the Hubble radius).
Deep inside this limit, the primordial spectra are given by
\begin{equation}
{\cal P}_{\cal R}(k) \simeq - \frac{H^2}{\pi m_P^2 (d \ln H/dN)}~,
\quad
{\cal P}_{h}(k) \simeq - \frac{16 H^2}{\pi m_P^2}~,
\label{slow-roll-spectra}
\end{equation}
where the right-hand sides are evaluated at Hubble crossing.  The
first-order expression of the scalar/tensor tilts $n_{S,T}$ and tilt
runnings $\alpha_{S,T}$ can be easily obtained by taking the
derivative of the above expressions, using the slow-roll approximation
$d/d \ln k \simeq d/dN$. The derivation of higher-order expressions is
more involved (see
e.g. \cite{Stewart:1993bc,Lidsey:1995np,Gong:2001he,Dodelson:2001sh,Schwarz:2001vv,Stewart:2001cd,Leach:2002ar,Hansen:2001eu,Caprini:2002jy,Liddle:2003py,Choe:2004zg,Habib:2005mh,Joy:2005ep,Kadota:2005hv,Casadio:2006wb,deOliveira:2005mf}).

\begin{figure*}[t]
\includegraphics{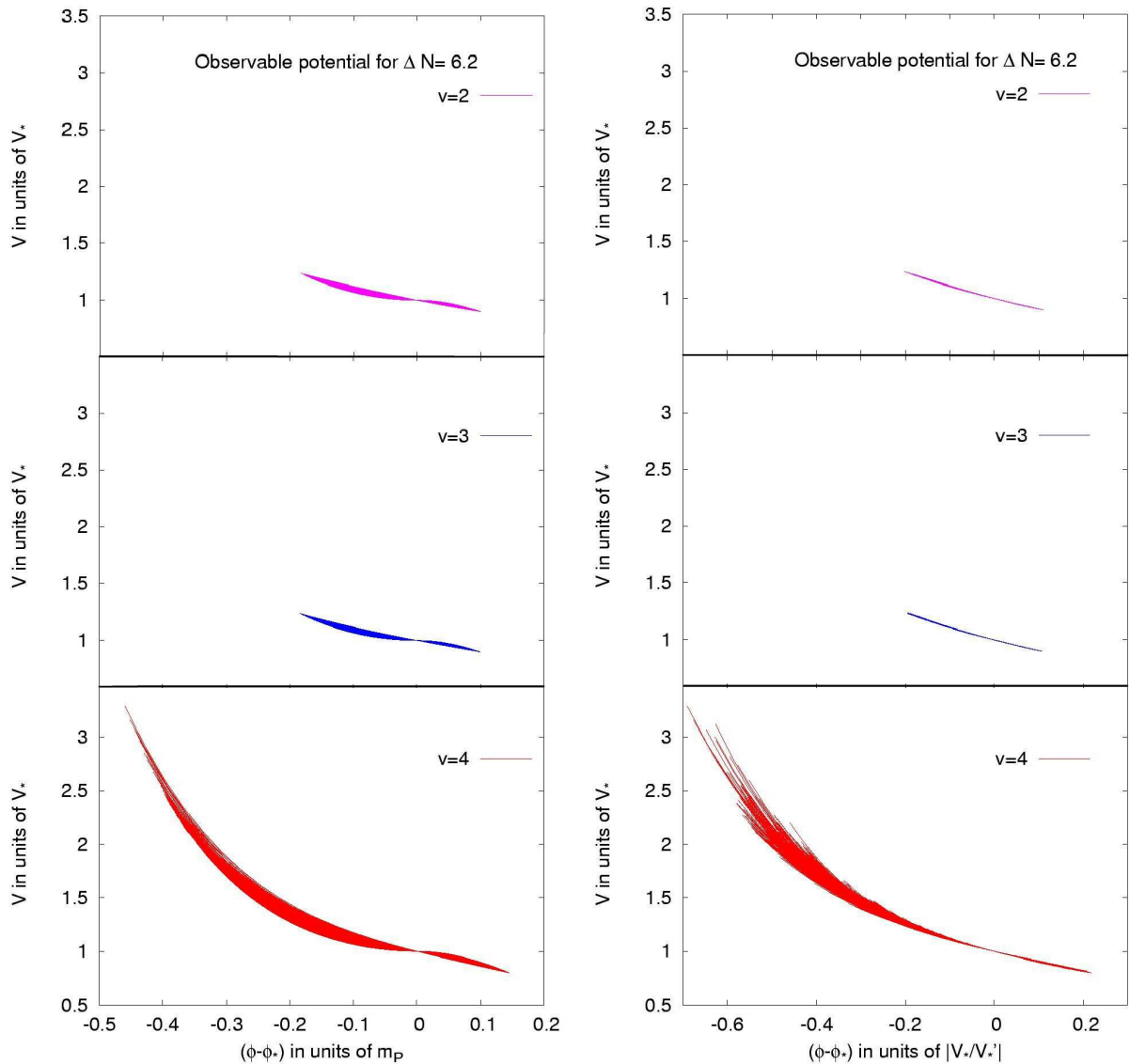}
\caption{\label{fig4} Observable region of the inflaton potential
allowed at the 95\% C.L. by the WMAP 3-year and Sloan Luminous Red
Galaxy survey (SDSS-LRG) data, for a Taylor expansion of the potential
at order $v=2$ (top), $v=3$ (middle) or $v=4$ (bottom), in the
vicinity of the pivot value $\phi_*$.  In all diagrams the potentials
are normalized to their value at the pivot scale $\phi_*$.  For
clarity, in the right diagrams, the field is expressed in units of
$|V_*/V'_*|$ instead of $m_P$, so all curves have by definition the
same slope in $\phi=\phi_*$.  In practice, these plots show the
superposition of 95\% of the potentials from our MCMC chains with the
best likelihood (after removal of the burn-in phase).
Each potential is plotted in a range $[\phi_1, \phi_2]$ corresponding
to Hubble exit for modes in the range $[k_1, k_2]$=$[2 \times 10^{-4},
0.1]~$Mpc$^{-1}$ which is most constrained by the data. This
corresponds roughly
to a history of 6.2 e-folds.  We only show here potentials with a
negative slope, but their image under the $\phi \longrightarrow -\phi$
symmetry are equivalent solutions. At first sight, on the top left
diagram, $v=2$ potentials seem to have a non-zero third derivative, but this
impression comes only from the superposition of many lines with varying
length and slope.  }
\end{figure*}
Current data clearly indicate that around the pivot scale at which the
amplitudes, tilts and runnings are defined (usually, the median scale
probed by the data), the tensor-to-scalar ratio is small and the
scalar tilt is close to one. This is sufficient for proving that the
two slow-roll conditions are well satisfied around the middle of the
observable e-fold range. However, depending on the inflaton potential,
higher derivatives $d^n \ln H / dN^n$ (with $n \ge 3$) could be large
near the pivot scale, leading to a sizable tilt running $\alpha_S$
and eventually to a situation in which slow-roll would hold only
marginally at the beginning and/or at the end of the observable e-fold
range. This explains why the two issues of large $|\alpha_S|$ and slow-roll
validity are closely related (as recently emphasized in
\cite{Makarov:2005uh}).

Since models with a large running imply that the two slow-roll
conditions become nearly saturated near the ends of the observable
potential range, a naive extrapolation would suggest that they cannot
sustain inflation for much more than the observable $\delta N \sim 10$
e-folds.  However, it is always conceptually easy to extrapolate the
potential in order to get the necessary 50 or 60 inflationary e-folds
after our observable universe has left the Hubble radius, or to obtain
arbitrary long inflation before that time.  Potentials designed in
that way might not have simple analytical expressions. This should not
be a major concern e.g. for physicists trying to derive inflation from
string theory, in which the {\it landscape} designed by the
multi-dimensional scalar potential can be very complicated, leading
{\it a priori} to any possible shape for the effective potential of
the degree of freedom driving inflation. However, it is clear that
models inducing large running are not as simple and minimalistic as
those with negligible running. But since they are not excluded, they
should still be considered in conservative works such as the present
one.

In section \ref{sec_spectrum}, we will follow a conventional approach
and fit directly the Taylor-expanded primordial spectra to the
data. Like most other authors, we will conclude that: (i) the data
provides absolutely no indication for $\alpha_S \neq 0$, and (ii)
given the current precision of the data, a large running is
nevertheless still allowed. We will show that similar conclusions also
apply to the running of the running $\beta_S$.

In section \ref{sec_potential}, which contains our main original
results, we will fit directly the Taylor-expanded scalar potential of
the inflaton to the data. Our reconstructed potentials are displayed
in Fig.~\ref{fig4}.  Unless we impose a ``no-running theoretical
prior'' (i.e., the prejudice that inflation is deep inside the
slow-roll regime), our potentials will freely explore the region in
parameter space where the running (and eventually the running of the
running) are as large as found in section \ref{sec_spectrum}. So, for
self-consistency, we must forget about the slow-roll formalism and
compute the exact primordial spectra numerically (as
Ref.~\cite{Martin:2006rs} did for various specific expressions of the
potentials). In a very nice work, Ref.~\cite{Makarov:2005uh} gave a
few examples of scalar potentials leading to the largest $|\alpha_S|$
values allowed by the data, and showed that even in these cases the
second-order slow-roll formalism, although inaccurate, remains a
reasonable approximation. In the present systematic analysis, which
explores the full parameter space of smooth inflationary potentials,
we will see that this conclusion does not apply in all cases.

\section{Fitting the primordial spectrum}
\label{sec_spectrum}

\begin{figure*}[]
\includegraphics[width=12cm]{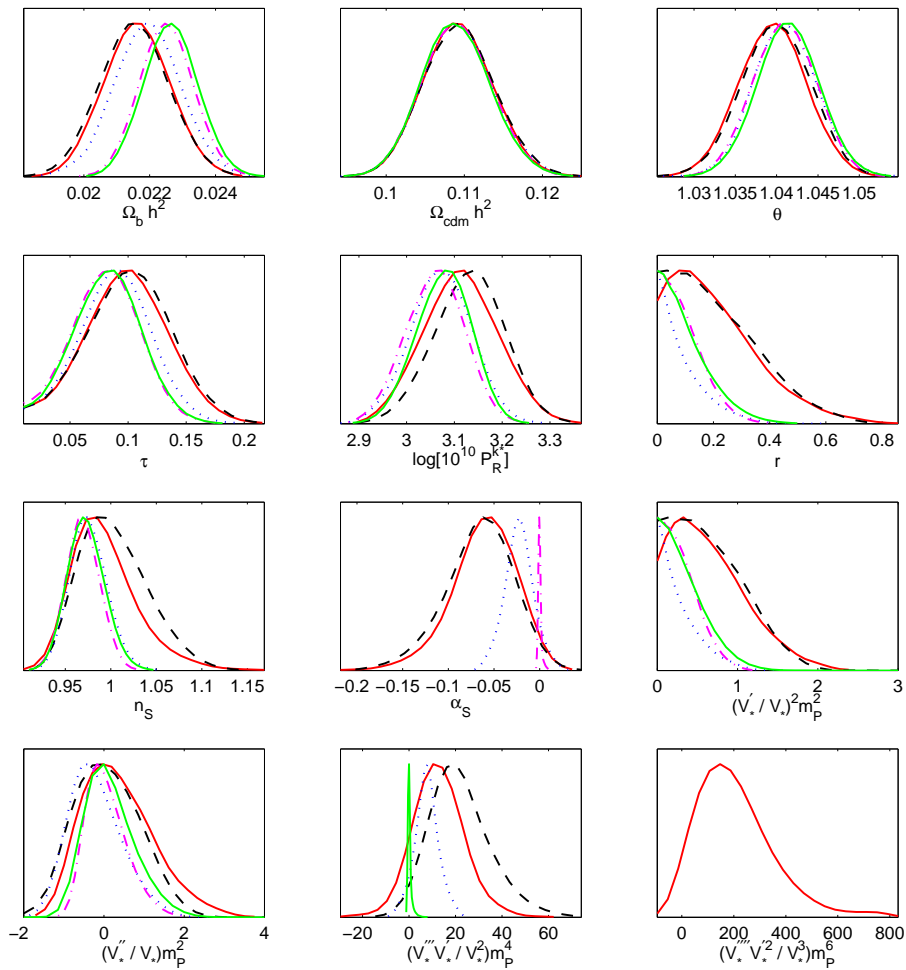}
\caption{\label{fig1} Probability distribution of cosmological and
inflationary parameters for the models of section \ref{sec_spectrum}:
$p=2$ (green/light solid), $p=3$ (black dashed), and the models of section
\ref{sec_potential}: $v=2$ (magenta dot-dashed), $v=3$ (blue dotted), 
$v=4$ (red/dark solid).
For the runs of section \ref{sec_spectrum}, the free parameters (with
flat priors) are the first eight parameters; the corresponding
probability for the potential parameters are derived from second-order
slow-roll formulae (involving $V_*$ to $V'''_*$, so the inferred value
of $V''''_*$ remains undetermined). Instead, for the runs of section
\ref{sec_potential}, the free parameters are the first five and the last four;
the amplitude parameter in the fifth plot is then defined as
$\ln \! \left[10^{10} \frac{128 \pi V^3_*}{3 V'^2_* m_P^6}\right]$;
the corresponding $r$, $n_S$ and $\alpha_S$ are derived from the exact
primordial spectra. The data consists of the WMAP 3-year
results~\cite{Spergel:2006hy,Page:2006hz,Hinshaw:2006ia,Jarosik:2006ib}
and the SDSS LRG spectrum~\cite{Tegmark:2006az}.}
\end{figure*}
{\it Primordial spectrum parametrization.}  The usual way of
testing inflationary models without making too many assumptions on the
inflaton potential is to fit some smooth scalar/tensor primordial
spectra, parametrized as a Taylor expansion of $\ln {\cal P}$ with
respect to $\ln k$,
\begin{equation}
\ln \frac{{\cal P}_{{\cal R}}(k)}{{\cal P}_{{\cal R}}(k_*)}
= (n_{S}-1) \ln \frac{k}{k_*}
+ \frac{\alpha_{S}}{2} \ln^2 \frac{k}{k_*}
+ \frac{\beta_{S}}{6} \ln^3 \frac{k}{k_*} ...,
\end{equation}
and the same holds for ${\cal P}_{h}$ as a function of $n_T$, $\alpha_T$
and $\beta_T$.
In single-field inflation, the coefficients of the
scalar and tensor spectra are related through the
approximate self-consistency relation
\begin{equation}
\frac{d \ln {\cal P}_{h} (k)}{d \ln k} \simeq \frac{1}{8} 
\frac{{\cal P}_{h} (k)}{{\cal P}_{\cal R} (k)} \label{slow-roll-self-consistency}
\end{equation}
which follows trivially from Eq.~(\ref{slow-roll-spectra}) and becomes
exact deep in the slow-roll limit. The sensitivity of current data to
gravitational waves is very low, with loose constraints on the {\it
shape} of ${\cal P}_{h}$.  So, even if the slow-roll formalism might become
inaccurate in some cases, the
data can be fitted assuming that Eq.~(\ref{slow-roll-self-consistency})
is exact.  In other words, for practical purposes, we can safely use
the hierarchy of relations derived from
Eq.(\ref{slow-roll-self-consistency}),
\begin{equation}
n_T=-r/8, \qquad \alpha_T=n_T [n_T-n_S+1], \qquad {\rm etc.}, 
\end{equation}
where $r \equiv {\cal P}_{h} (k_*) / {\cal P}_{\cal R} (k_*)$.  So, if
we decide to Taylor expand the scalar spectrum with $p$ independent
coefficients, the total number of free inflationary parameters in the
problem is $p$+1, including the tensor-to-scalar amplitude ratio at
the pivot scale.
\begin{figure}[]
\includegraphics[width=8.5cm]{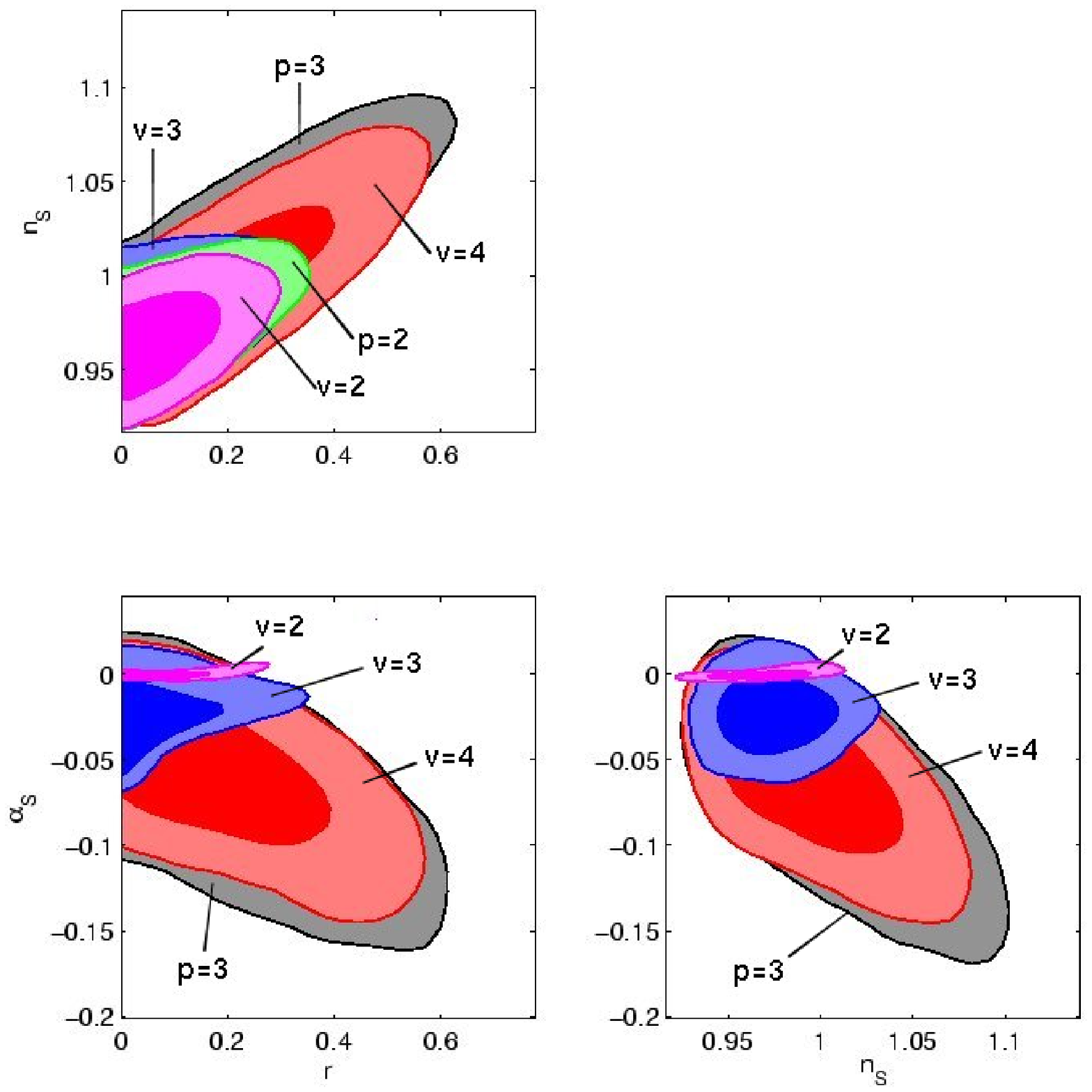}
\caption{\label{fig3} Two-dimensional 68\% and 95\% confidence level
contours based on WMAP 3-year and the SDSS LRG spectrum 
for the parameters describing the primordial spectra,
obtained directly from the MCMC in the case of models
$p=2$ (green) and $p=3$ (black), or derived form the exact spectra for models
$v=2$ (magenta), $v=3$ (blue), $v=4$ (red).}
\end{figure}

In principle, $p$ should be chosen according to Occam's razor: when
increasing $p$ does not improve sufficiently the goodness-of-fit, it
is time to stop. In a Bayesian analysis, this question is addressed by
the computation of the Bayesian
evidence~\cite{Beltran:2005xd,Trotta:2005ar,Kunz:2006mc,Parkinson:2006ku,Pahud:2006kv,Liddle:2006tc,Liddle:2007fy}.
However, when the evidence does not vary significantly as a function
of $p$, the decision of stopping the expansion remains a personal
choice to some extent, and more conservative works should consider
higher $p$ values.

The issue of varying $p$ is important in two respects: first, one
needs to know how many independent informations the data is providing,
i.e., how smooth/complicated the inflaton potential needs to be (for
addressing this issue, one could also perform a principal component
analysis~\cite{Leach:2005av}); second, it is useful to know whether
the bounds on a given cosmological/inflationary parameter $\theta_i$
are independent of $p$, or subject to variations when $p$ increases,
due to the appearance of new parameter degeneracies.

\vspace{0.5cm}

{\it Results.}  In order to address these two points, we performed
some global parameter fits using the public code
{\sc cosmomc}~\cite{Lewis:2002ah}, with $p$ varying from two (scalar
amplitude and tilt) to four (including the tilt running, as well as
the running of the running). Our results are summarized in
Table~\ref{table1}.
\begin{table}[]
\begin{center}
\begin{tabular}{l|ccc}
Parameter & $p=2$ & $p=3$ & $p=4$ \\
\hline
\hline
$\Omega_b h^2$ & 
$0.0226 \pm 0.001$ & $0.022 \pm 0.001$ & $0.021 \pm 0.001$ \\
$\Omega_{cdm} h^2$ & 
$0.109 \pm 0.004$ & $0.109 \pm 0.004$ & $0.109 \pm 0.005$ \\
$\theta$ & 
$1.041 \pm 0.004$ & $1.040 \pm 0.004$ & $1.041 \pm 0.004$ \\
$\tau$ & 
$0.08 \pm 0.01$ & $0.10 \pm 0.02$ & $0.10 \pm 0.02$ \\
$\ln[10^{10} {\cal P}_{\cal R}^{k_*}]$ &
 $3.08 \pm 0.06$ & $3.13 \pm 0.07$ & $3.13 \pm 0.07$ \\
$r$ & 
$<0.13$ & $<0.3$ & $<0.3$ \\
$n_S$ & 
$0.97 \pm 0.02$ & $1.00 \pm 0.04$ & $1.03 \pm 0.05$ \\
$\alpha_S$ & 
0 & $-0.07 \pm 0.04$ & $-0.07 \pm 0.04$ \\
$\beta_S$ & 
0 & 0 & $-0.04 \pm 0.04$ \\
\hline
$- \ln {\cal L}_{\rm max}$ & 2688.3 & 2687.1 & 2686.5 \\
$E$ & 1 & 1.2 & 1.4 \\
\end{tabular}
\end{center}
\caption{\label{table1} Bayesian 68\% confidence limits for
$\Lambda$CDM inflationary models with $p=2,3,4$ coefficients in the
logarithmic Taylor expansion of the scalar primordial spectrum. The
last lines show the maximum likelihood value and the Bayesian
Evidence (relative to that of $p=2$).  The data consists in the WMAP
3-year
results~\cite{Spergel:2006hy,Page:2006hz,Hinshaw:2006ia,Jarosik:2006ib}
and the SDSS-LRG spectrum~\cite{Tegmark:2006az}, as implemented in
{\sc cosmomc}~\cite{Lewis:2002ah}.  }
\end{table}
The relative Bayesian evidence of each model can be easily computed,
since the models are nested inside each other
\cite{Trotta:2005ar}. However, this calculation forces us to choose
some explicit Bayesian priors for $\alpha_S$ and $\beta_S$.
Pushing inflation to its limits, we notice that the second slow-roll
parameter can in principle vary between plus and minus one, so
the scalar tilt could take any value between zero and two.  Extreme
runnings could be observed in ad-hoc inflationary models such that
during the observable e-folds, corresponding to four decades in $k$
space, $n_S$ evolves from 0 to 2 or vice-versa. So, the $\alpha_S$
prior can be chosen to be a top-hat centered on zero with $\Delta
\alpha_S=4/\ln(10^4)\sim0.4$. Similarly, extreme values of $\beta_S$
correspond to $n_S$ passing through the sequence 0-2-0 or 2-0-2. This
leads to a prior width $\Delta \beta_S=16/\ln^2(10^4) \sim 0.2$.
%
%
With such priors, the Bayesian evidence $E$ increases by a factor
\begin{equation}
\frac{E_{p=3}}{E_{p=2}} = \left[ {\cal P}(\alpha_S=0) \, \Delta
\alpha_s \right]^{-1} = [2.1 \times 0.4]^{-1} = 1.2
\end{equation} when 
$\alpha_S$ is added, and again by 
\begin{equation}
\frac{E_{p=4}}{E_{p=3}}=  \left[ {\cal P}(\beta_S=0) \, \Delta
\beta_s \right]^{-1} = [6.2 \times 0.2]^{-1} = 1.2
\end{equation} 
when $\beta_S$ is introduced. These numbers are too close to one for
drawing definite conclusions: the extra parameters are neither
required, neither disfavored by Occam's Razor.

Table~\ref{table1} shows that adding $\alpha_S$ has a small impact on
the probability distribution of $\Omega_bh^2$, $\tau$, $r$, ${\cal
P}_{\cal R}(k_*)$ and $n_S$, as found in previous works.  However, it
is reassuring to note that adding $\beta_S$ leave all bounds perfectly
stable, except for a small shift to higher $n_s$ values. This
suggests that including a few higher derivatives beyond $\alpha_S$
does not open new parameter degeneracies (this conclusion would
probably break if the number of free parameters becomes much
larger). Figs.~\ref{fig1},~\ref{fig3} show the likelihood distribution
of each parameter as well as some two-dimensional
confidence regions for models $p=2$ (green lines) and $p=3$ (black lines).

\vspace{0.5cm}

{\it Impact of small CMB multipoles.} The smallest multipoles in the
CMB temperature and polarization maps are still controversial. In WMAP
data (as well as in previous COBE data), the temperature quadrupoles
and octopoles are surprisingly small, while their orientations seem to
be correlated (between each other and with the ecliptic plane). Many
authors have been investigating possible foregrounds or systematics
which could affect these small multipoles (see
e.g.~\cite{Schwarz:2004gk,Copi:2005ff,Copi:2006tu} and references
therein). So, it is
legitimate to study whether the
quadrupole and octopole data have a significant impact on our bounds for the
primordial spectrum parameters (a priori, these low temperature
multipoles could be partially responsible for the preferred negative
value of the tilt).  

We repeated the $p=3$ analysis after cutting the temperature and
polarization data at $l=2,3$. We found that all probabilities are
essentially unchanged, including that for running (the mean value only
moves from $-0.67$ to $-0.66$, which is not significant given the
precision of the runs).  We conclude that our results are independent
of the robustness of low mutipole data.\footnote{This result is
consistent with that of Ref.~\cite{Cline:2006db}, which shows that evidence for
running is more related to anomalies around $l \sim 40$.}

\vspace{0.5cm}

{\it Impact of extra CMB data.} There are also discussions about a
possible small mismatch in the amplitude of the third acoustic peak
probed on the one hand by WMAP, and on the other hand by
Boomerang~\cite{Jones:2005yb} or other small-scale CMB experiments.
We repeated the $p=2$ and $p=3$ analysis including extra data from
Boomerang~\cite{Jones:2005yb}, ACBAR~\cite{Kuo:2002ua} and
CBI~\cite{Sievers:2005gj}. The impact on inflationary parameter is
found to be very small, although in the $p=3$ case the bound on $r$
gets weaker by 20\% and the preferred value of $\alpha_S$ goes down by
the same amount (i.e., the case for negative running becomes slightly
stronger). Other bounds are essentially unchanged. In what follows, we
will not include these data sets anymore.

\vspace{0.5cm}

{\it Expectations for the inflaton potential.}  In the next section,
we will directly fit the inflaton potential $V(\phi)$, parametrized as
a Taylor expansion near the value $\phi_*$ corresponding to Hubble
crossing for the pivot scale $k_*$.  We expect that a global fit with
$V(\phi)$ expanded at order $v$ will provide the same qualitative
features as the previous power spectrum fit of order $p=v$:
\begin{itemize}
\item order $v=2$ ($V'''=0$)
should be sufficient for explaining the data,
and will not lead to significant running or deviation from
slow-roll.  Indeed, the smallness of $r$ and $|n_S-1|$ guarantees that
the two slow-roll conditions are well satisfied at least near $\phi_*$.
In addition, with $V'''=0$, they should remain well satisfied
on the edges of the spectrum, and no significant running can be generated.
\item order $v \geq 3$ ($V''''=0$) should not be required
by the data, but remains interesting since it will explore the
possibility of large running and shift the other parameter
distributions as in the previous $p=3$ case. The slow-roll parameters
could then become large near the edges, so it is 
necessary to compute the spectra numerically rather than
using any slow-roll approximation.
\end{itemize}
The results of the next section will confirm this expectation, and
prove that order $v=4$ is necessary for exploring the full range of
$\alpha_s$ probed by the $p=3$ run.

\section{Fitting the scalar potential}
\label{sec_potential}

{\it Computing the power spectra numerically.}  In order to fit
directly the inflaton potential, we wrote a new {\sc cosmomc} module
which computes the scalar and tensor primordial spectra exactly, for
any given function $V(\phi-\phi_*)$. This module can be downloaded
from the website 
\url{http://wwwlapp.in2p3.fr/~lesgourgues/inflation/}, and easily
implemented into {\sc cosmomc}.

In its present form,
our code is not designed for models with very strong deviations from
slow-roll. For such extreme models, a given function $V(\phi-\phi_*)$
would not lead to a unique set of primordial spectra ${\cal P}_{\cal
R}(k)$, ${\cal P}_{h}(k)$: the result would depend on the initial
conditions in phase space. We decide to limit ourselves to models such
that throughout the observable range, the field remains close to the
attractor solution for which $\dot{\phi}$ is a unique function
$\phi$. In this case, a given function $V(\phi-\phi_*)$ does lead to
unique primordial spectra, and we do not need to introduce an extra
parameter $\dot{\phi}_{\rm ini}$.
Since the goal of this paper is to test
inflationary potentials leading to smooth primordial spectra, this
restriction is sufficient. In particular, it enables to explore models
for which the running $\alpha_S$ is large, deviations from slow-roll
are significant, and analytical derivations of the spectra are
inaccurate.  However, our code cannot deal with the case in which
inflation starts just when our observable universe crosses the horizon
(for which $\dot{\phi}_{\rm ini}$ would be a crucial extra free parameter).

In {\sc cosmomc}, we fix once and for all the value of the pivot scale
$k_*=0.01~$Mpc$^{-1}$. Then, for each function $V(\phi-\phi_*)$ passed by
{\sc cosmomc}, our code computes the spectra ${\cal P}_{\cal R}(k)$,
${\cal P}_{h}(k)$ within the range $[k_{\rm min}, k_{\rm
max}]=[5\times 10^{-6},5]~$Mpc$^{-1}$ needed by {\sc camb}, imposing that
$aH=k_*$ when $\phi=\phi_*$. So, the code finds the attractor solution
around $\phi=\phi_*$, computes $H_*$ and normalizes the scale factor
so that $a_*=k_*/H_*$. Then, each mode is integrated numerically for
$k/aH$ varying between two adjustable ratios: here, $50$ and
$1/50$. So, the earliest (resp. latest) time considered in the code is
that when $k_{\rm min}/aH=50$ (resp.  $k_{\rm max}/aH=1/50$),
%
which in the attractor solution uniquely determines extreme values of
$(\phi-\phi_*)$ according to some potential. In the code this is
translated to demanding that $aH$ grows according to the
aforementioned ratios: by $50 k_*/ k_{\rm min}$ before $\phi=\phi_*$, and
by $50 k_{\rm max}/k_*$ afterwards.
%
Hence, one of the preliminary tasks of the code is to find the
earliest time. If by then, a unique attractor solution for the
background field cannot be found within a given accuracy (10\% for
$\dot{\phi}_{\rm ini}$), the model is rejected.  So, we implicitly
assume that inflation starts at least a few e-folds before the present
Hubble scale exits the horizon.  In addition, we impose a positive,
monotonic potential and an accelerating scale factor during the period
of interest.  This prescription discards any models with a bump in the
inflaton potential or a short disruption of inflation, that could
produce sharp features in the power spectra.

%
As a result of the chosen method, the potential is slightly
extrapolated beyond the observable window, in order to reach the
mentioned conditions for the beginning and ending of the numerical
integration. Although this seems to be in contradiction with the
purpose of this paper, i.e. to probe only the observable potential,
this extrapolation cannot be avoided if we want to keep the number of
free parameters as small as possible. Note that the range of
extrapolation is still very small in comparison with an extrapolation
over the full duration of inflation after the observable modes have
exited the Hubble radius.
%

%
In this approach we need not make any assumption about reheating and
the duration of the radiation era. As explained in
Ref.~\cite{Ringeval:2007am}, the evolution during reheating determines
the redshift $z_*$ at which presently observed perturbations left the Hubble
radius during inflation. Probing only the observable window of
perturbations, we are allowed to let the subsequent evolution of
inflation and reheating, hence the number of e-folds and thereby
the redshift $z_*$, be unknown.
%

\vspace{0.5cm}

{\it Parametrization.}  The inflaton potential is Taylor expanded up
to a fixed order, and we let {\sc cosmomc} probe different values for
the derivatives of the inflaton potential at the pivot scale. Since
Monte Carlo Markov Chains (MCMC) converge faster if the probed
parameters are nearly Gaussian distributed, in fact we recombine to
potential parameters in such a way as to probe nearly Gaussian
combinations. These combinations are inspired by the slow-roll
expression of the spectral parameters (${\cal P}_{\cal R}(k_*)$,
$n_S$, $\alpha_S$ and $r$) as a function of the potential. We use an
amplitude parameter $ \frac{128 \pi}{3} \frac{V^3_*}{V'^2_* m_P^6}$,
which is equal to ${\cal P}_{\cal R}(k_*)$ at leading order in a
slow-roll expansion.  The other spectral parameters consist of linear
combinations of $(V_*'/V_*)^2$, $V''_*/V$, $(V'''_*/V_*)(V_*'/V_*)$
and $(V''''_*/V_*)(V'_*/V_*)^2$. Hence it is most likely to find
nearly Gaussian shapes for these products in stead of the sole
potential derivatives. For the actual expressions for the spectral
parameters in terms of the inflaton potential, we refer the reader e.g. to
section IV of~\cite{Leach:2002ar}.

\begin{table}[t]
\begin{center}
\begin{tabular}{l|ccc}
Parameter & $v=2$ & $v=3$ & $v=4$ \\
\hline
\hline
$\Omega_b h^2$ & 
$0.022 \pm 0.001$ & $0.022 \pm 0.001$ & $0.022 \pm 0.001$ \\
$\Omega_{cdm} h^2$ & 
$0.109 \pm 0.004$ & $0.109 \pm 0.004$ & $0.109 \pm 0.004$ \\
$\theta$ & 
$1.041 \pm 0.004$ & $1.041 \pm 0.004$ & $1.040 \pm 0.004$ \\
$\tau$ & 
$0.08 \pm 0.03$ & $0.09 \pm 0.03$ & $0.10 \pm 0.03$ \\
$\ln \! \left[\frac{128 \pi 10^{10} V^3_*}{3 V'^2_* m_P^6}\right]$ &
 $3.06 \pm 0.06$ & $3.07 \pm 0.06$ & $3.11 \pm 0.08$ \\
$\left(\frac{V'_*}{V_*}\right)^2 \!\!\! m_P^2$ & 
$<0.4$ & $<0.4$ & $<0.8$ \\
$\frac{V''_*}{V_*} m_P^2$ & 
$0.1 \pm 0.5$ & $-0.2 \pm 0.6$ & $0.4 \pm 0.9$ \\
$\frac{V'''_*}{V_*} \frac{V'_*}{V_*} m_P^4$ & 
0 & $8 \pm 5$ & $13 \pm 11$ \\
$\frac{V''''_*}{V_*} \left(\frac{V'_*}{V_*} \right)^2 \!\!\! m_P^6$ & 
0 & 0 & $200 \pm 150$ \\
\hline
$- \ln {\cal L}_{\rm max}$ & 2688.3 & 2687.2 & 2687.2 \\
\end{tabular}
\end{center}
\caption{\label{table2} Bayesian 68\% confidence limits for
$\Lambda$CDM inflationary models with a Taylor expansion of the
inflaton potential at order $v=2,3,4$ (with the primordial spectra
computed numerically). The last line shows the maximum
likelihood value.  The data consists of the WMAP 3-year
results~\cite{Spergel:2006hy,Page:2006hz,Hinshaw:2006ia,Jarosik:2006ib}
and the SDSS LRG spectrum~\cite{Tegmark:2006az}, as implemented in
{\sc cosmomc}~\cite{Lewis:2002ah}.  }
\end{table}
In order to compare the results for the runs with $v=2,3,4$ with those
of the previous section, we calculate the spectral parameters of each
model numerically.  The other way around, we also invert the slow-roll
expansion in order to compare the $v=2,3,4$ and $p=2,3$ models in
potential-derivative space. Defining $\epsilon_0\equiv H(N_I)/H(N)$
and $\epsilon_{n+1}\equiv\frac{d \ln \left|\epsilon_{n}\right|}{dN}$,
where $N=\ln \frac{a}{a_i}$ and $H_i$ is some initial value of the
Hubble factor, the inversion is given by
\begin{eqnarray}
\epsilon_1&=&\frac{r}{16}+\frac{C_1}{16}
   \left(\frac{r^2}{8}+(n_S-1)
   r\right)\nonumber\\
&&+\mathcal{O}\left(r^3,\left(n_S-1\right)^3,\alpha_S^3\right),\\
\epsilon_2&=&
-(ns-1)+C_1\alpha_S-\frac{r}{8}-\frac{r}{8}(ns-1)\left(C_1-\frac{3}{2}\right)\nonumber\\
&&-\left(\frac{r}{8}\right)^2(C_1-1)+\mathcal{O}\left(r^3,\left(n_S-1\right)^3,\alpha_S^3\right),\\
\epsilon_2\epsilon_3&=&\frac{1}{8}
   \left(\frac{r^2}{8}+(n_S-1) r-8 \alpha_S
   \right)\nonumber\\
&&+\mathcal{O}\left(r^3,\left(n_S-1\right)^3,\alpha_S^3\right),
\end{eqnarray}
where $C_1 =\gamma_E+\ln 2 -2\simeq -0.7296$.  The value of the
potential and its derivatives can be expressed exactly in terms of the
slow-roll parameters, which are listed up to the second derivative in
\cite{Leach:2002ar}. The third derivative reads
\begin{equation}
V'''=\frac{12 m_p^2 H^2  \sqrt{\pi } }{\sqrt{\epsilon_1}} \left(2
   \epsilon_1^2-\frac{3 \epsilon_2
   \epsilon_1}{2}+\frac{\epsilon_2\epsilon_3}{4}\right).
\end{equation}
The fourth derivative of the inflaton potential would be of a higher
order in the slow-roll expansion. 

\vspace{0.5cm}

\begin{figure*}[t]
\includegraphics[width=12cm]{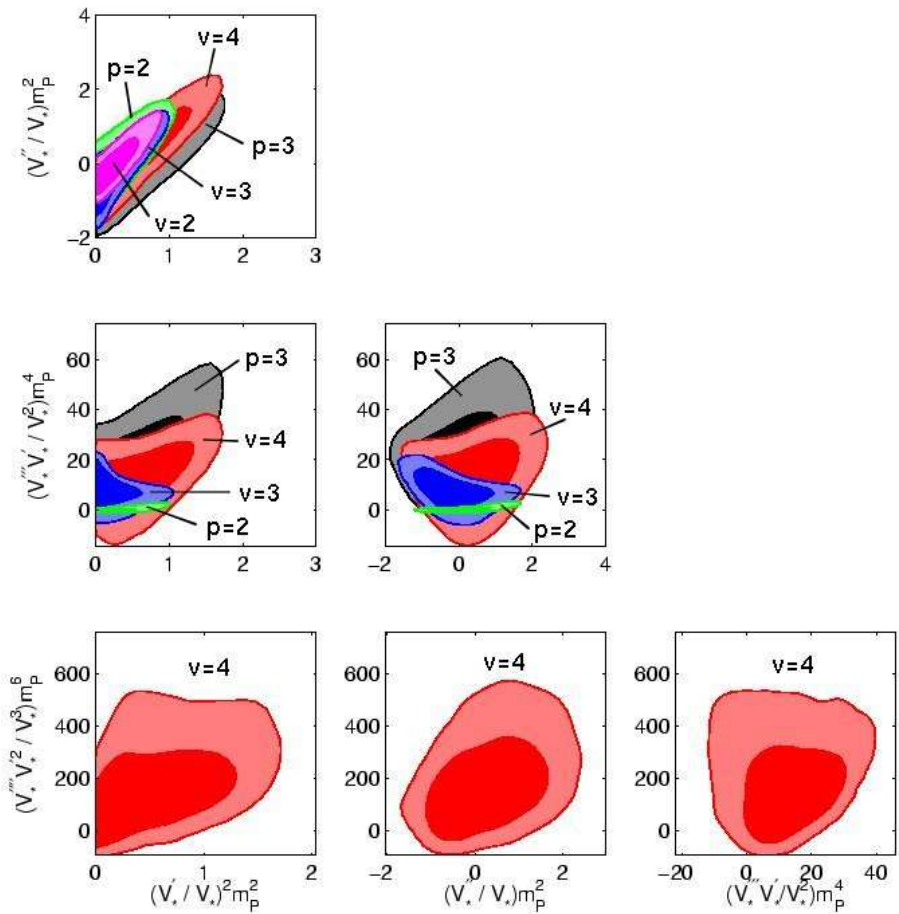}
\caption{\label{fig2} Two-dimensional 68\% and 95\% confidence level
contours based on WMAP 3-year and the SDSS LRG spectrum, for
the parameters describing the inflaton potential,
obtained directly from the MCMC in the case of models 
$v=2$ (magenta), $v=3$ (blue), $v=4$ (red), or derived
from second-order formulas for models $p=2$ (green), $p=3$ (black).}
\end{figure*}
\begin{figure*}[t]
\includegraphics[angle=-90,width=8.5cm]{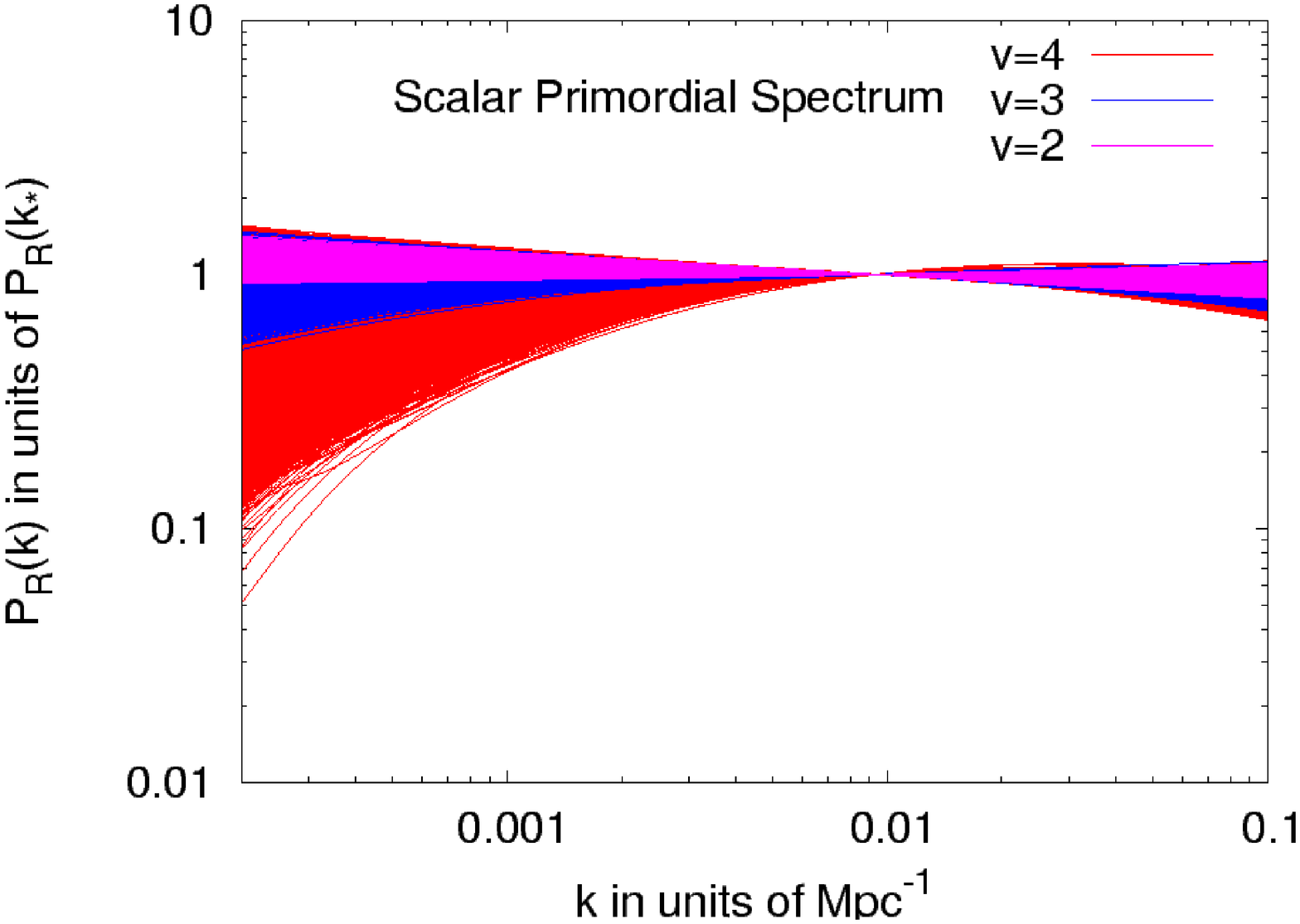}
\includegraphics[angle=-90,width=8.5cm]{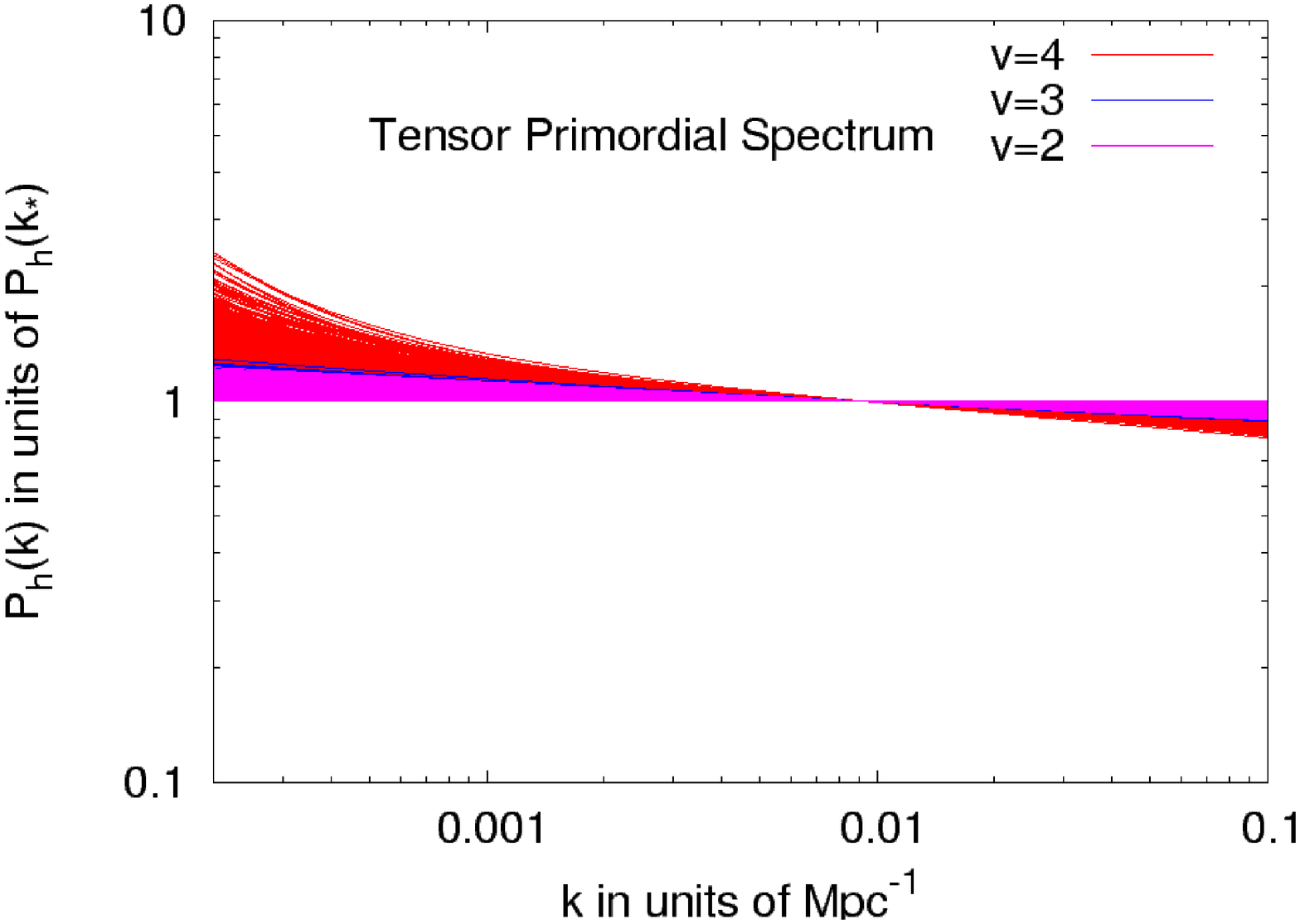}
\caption{\label{fig5} The primordial spectrum for scalar perturbations
(left) and tensor perturbations (right) allowed at the 95\% C.L. by
WMAP 3-year and the SDSS LRG data, for a Taylor expansion of the
inflaton potential at order $v=2$ (magenta/light), $v=3$ (blue/dark) or $v=4$
(red/medium). In practice, this plot shows the superposition of 95\% of the
spectra from our MCMC chains with the best likelihood (after removal
of the burn-in phase).  All these spectra are computed numerically,
rescaled to one at the pivot value $k_*$, and displayed in the range
which is most constrained by our data set. Note that the shapes of the
two spectra are related to each other: so, the
tensor spectrum is constrained through that of the scalar spectrum,
and not directly by the data, which does not have the required
sensitivity.}
\end{figure*}
\begin{figure*}[t]
\includegraphics[angle=-90,width=6cm]{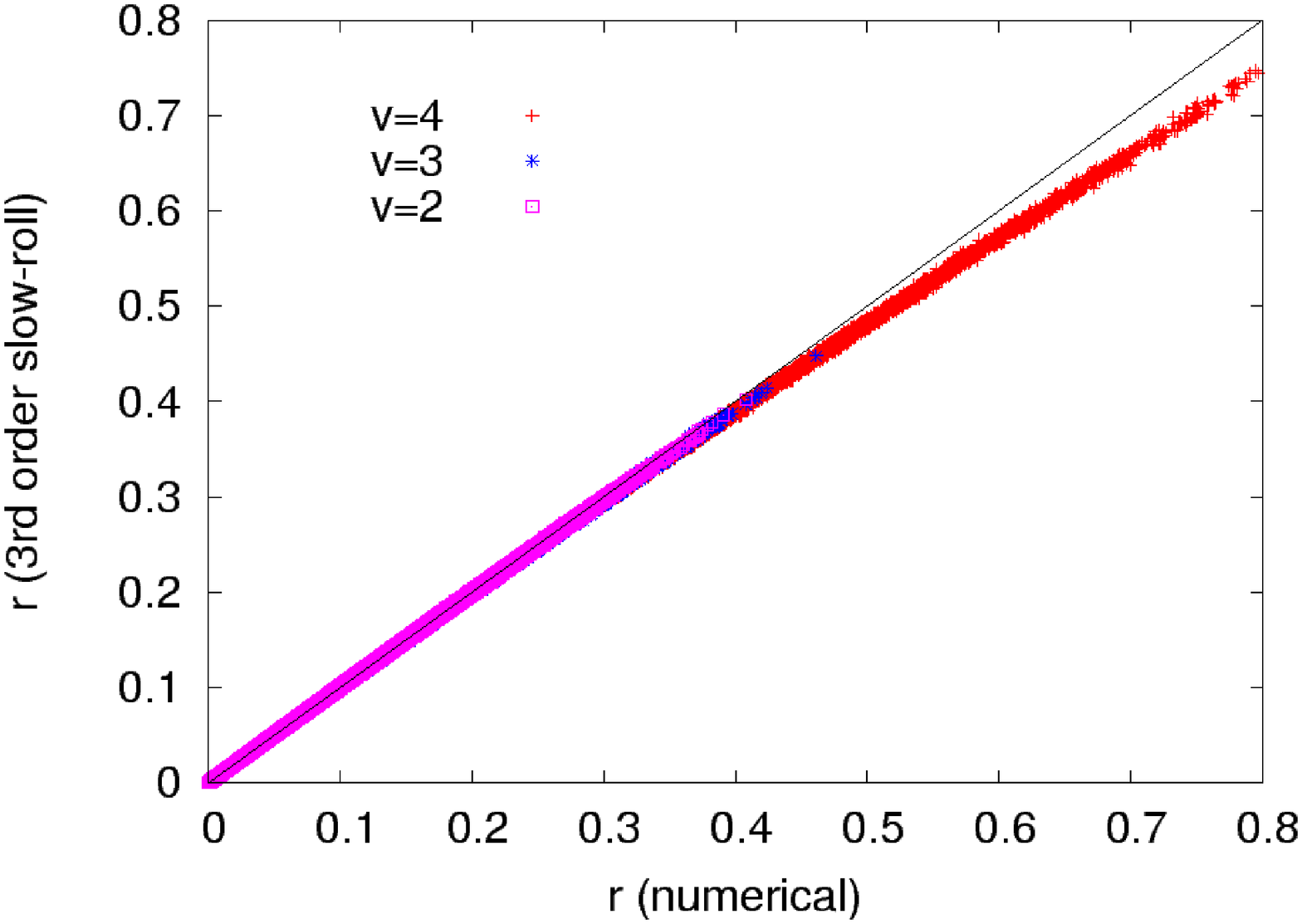}
\includegraphics[angle=-90,width=6cm]{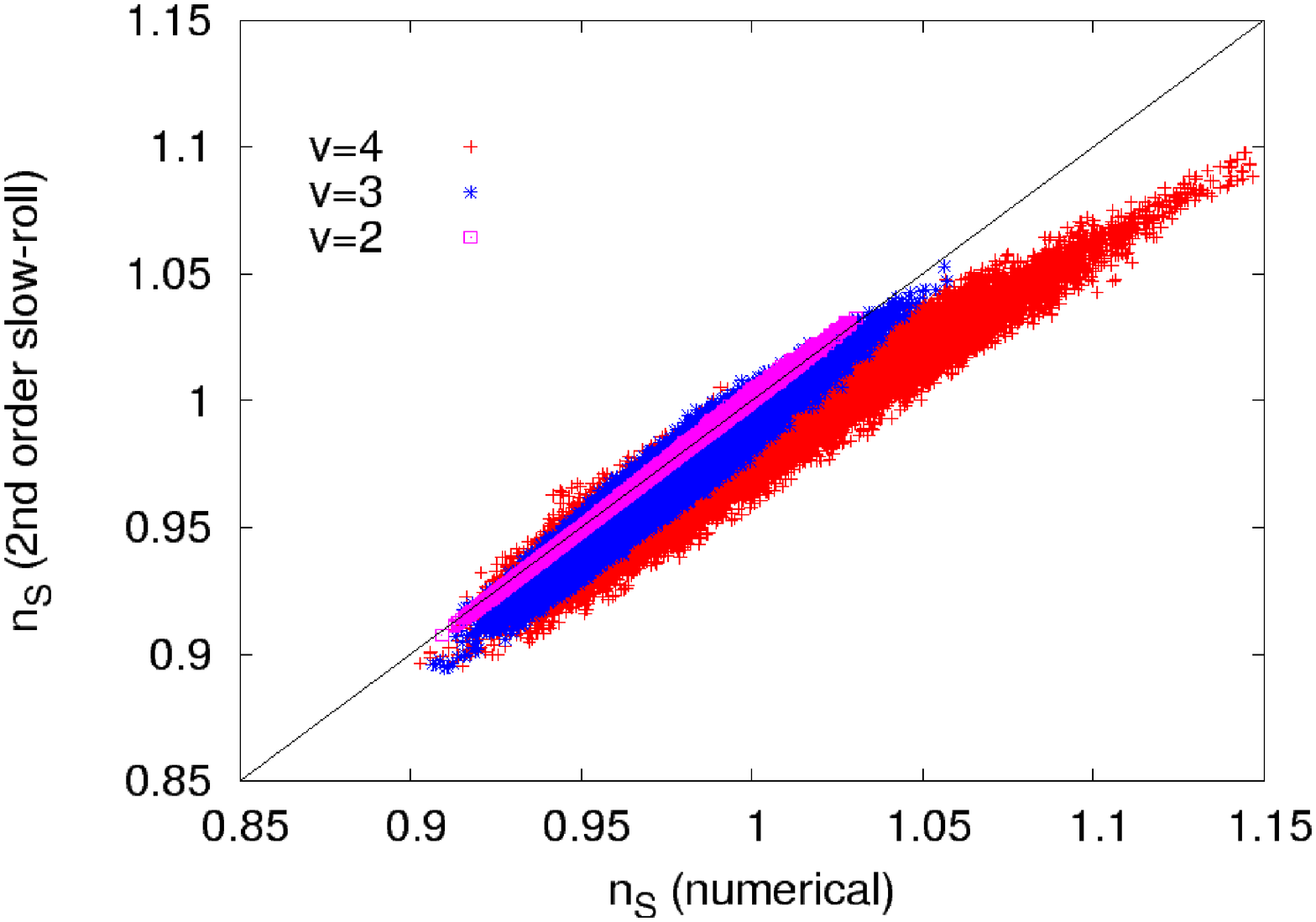}\\
\includegraphics[angle=-90,width=6cm]{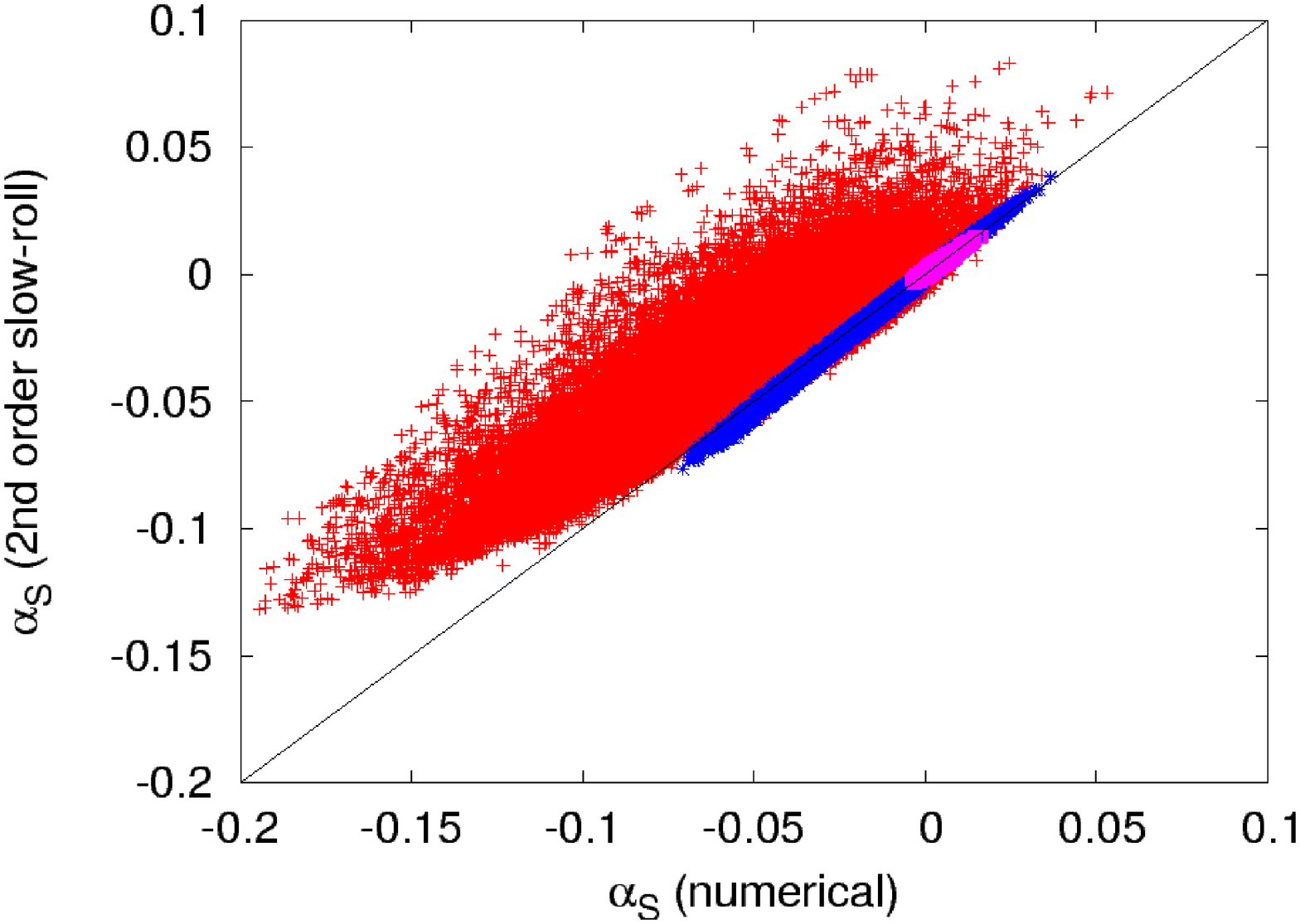}
\includegraphics[angle=-90,width=6cm]{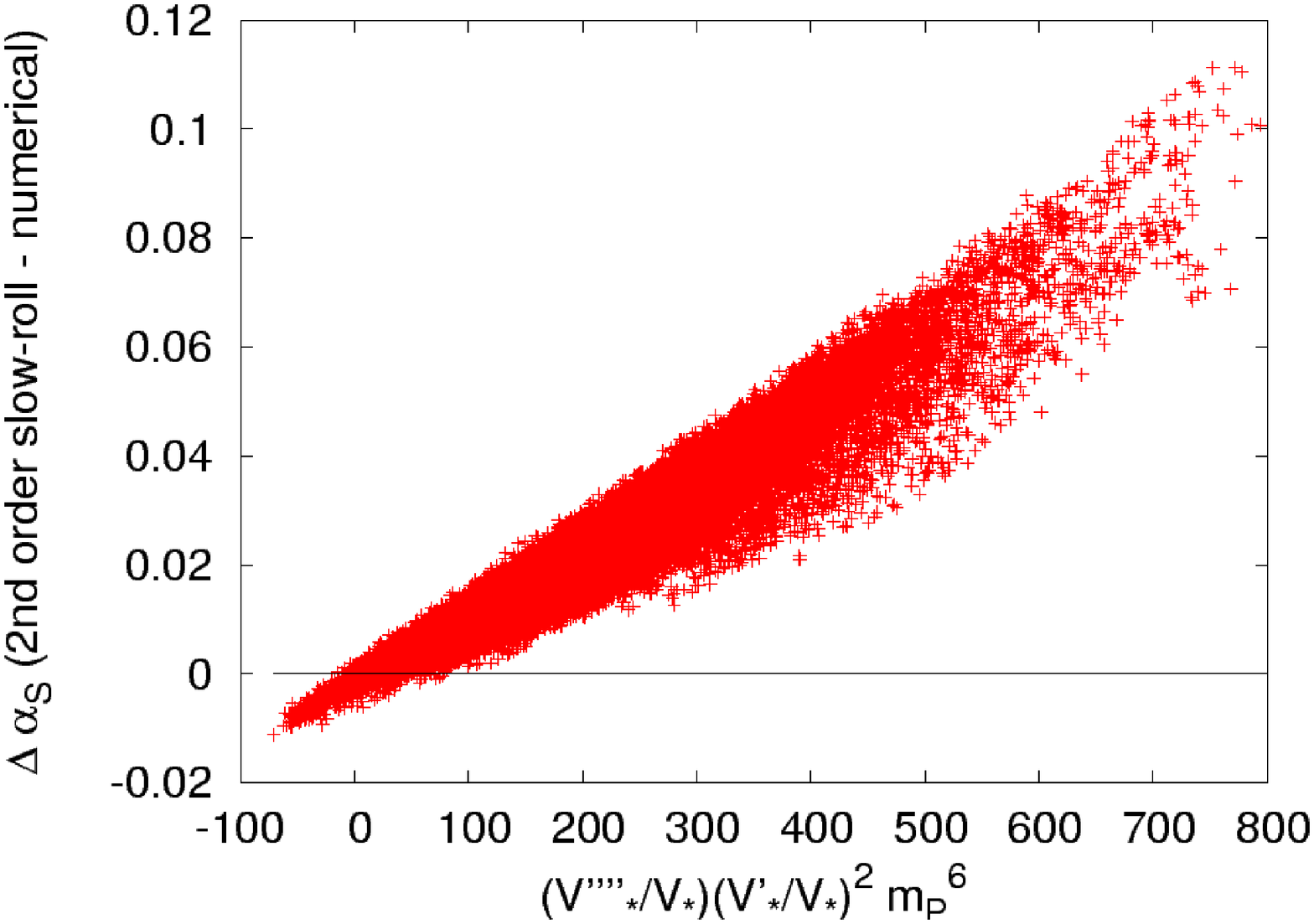}
\caption{\label{fig6} Precision test of the second-order slow-roll
expansion for models allowed at the 95\% C.L. by WMAP 3-year and the
SDSS LRG data, based on a Taylor expansion of the inflaton potential
at order $v=2$ (magenta/light), $v=3$ (blue/dark) or $v=4$ (red/medium).  For each
model, we plot the spectral parameter $r$ (top left), $n_S$ (top
right) and $\alpha_S$ (bottom left) computed at the pivot scale with
two methods: by deriving the primordial spectrum computed numerically
(horizontal axis), or with the second-order slow-roll formalism. We
checked that the scattering of the point away from the y=x axis
reflects inaccuracies in the second-order slow-roll formalism rather
than in our code.  The bottom right plot shows the inaccuracy in the
running as a function of $V''''_* {V'}_*^2/V_*^3$ for the $v=4$ model.}
\end{figure*}
{\it Results.}  The allowed ranges, parameter likelihoods and
two-dimensional contours from all our runs are summarized respectively in
Table~\ref{table2}, Fig.~\ref{fig1} and Figs.~\ref{fig3},~\ref{fig2}.
The allowed shape of primordial scalar and tensor spectra is shown in
Figs.~\ref{fig5}.

First we ran a chain for the model at order $v=2$. As expected, the
results confirm those obtained fitting the spectral parameters up to
order $p=2$, which can be seen in Figs.~\ref{fig1}~and~\ref{fig2} and
the upper left chart in Fig.~\ref{fig3}, by comparing the magenta and
green lines. This can be translated into the statement that fixing the
running of the tilt to zero is almost equivalent to fixing third and
higher derivatives of the potential to zero. The resulting bounds can
be read from Table~\ref{table2}, and the correlation between
$V'_*/V_*$ and $V''_*/V_*$ is well accounted by the relation
\begin{equation}
m_P^2 \left[2.2\left(\frac{V'_*}{V_*}\right)^2 - \frac{V''_*}{V_*} \right]
= 0.6 \pm 0.2~. \qquad (68\% {\rm C.L.})
\end{equation}
Note that the numerically calculated running in the models with $v=2$
is not strictly zero but allows a very small region of nonzero
running. Similarly, the derived bounds from the models with $p=2$ on
the potential parameters allow for very small regions of nonzero
second and third derivative. This merely reflects the expansions in
different parameterizations than an indication for running (or nonzero
higher derivatives of the potential).

Including a third derivative does allow for models to have a more
significant running, which is clearly visible Fig.~\ref{fig5}. Yet, as
seen in Figs.~\ref{fig1},~\ref{fig3}~and~\ref{fig2}, the models with
$v=3$ (blue line) do not explore the full range of parameters which is
indicated by the models with $p=3$ (black line), in particular for
$\alpha_S$, and do not show as much a sign of degeneracy between
$V'''_*V'_*/V_*^2$ and $V_*'^2$ as the derived potential derivatives
from the models with $p=3$, in Fig.~\ref{fig2}. The relation between
$V''$ and $V'^2$ remains almost unchanged. Inversely, in
Fig.~\ref{fig3} we see the same effect in spectral parameter space.

The remaining discrepancy between models $v=3$ and $p=3$ led us to
including the fourth derivative of the inflaton potential as a free
parameter, i.e. $v=4$. The resulting power spectra, shown in
Fig.~\ref{fig5}, show a larger negative running than in the $v=3$
case, even with significant running of the running on the largest
scales. In Fig.~\ref{fig1} we see that the model $v=4$ (red line) does
probe the same range of runnings of the tilt as allowed in the model
$p=3$. Looking at two-dimensional projections, we see that the $p=3$
and $v=4$ contours are closer to each other in spectral parameter
space (Fig.~\ref{fig3}) than in potential parameter space
(Fig.~\ref{fig2}): this reflects the inaccuracy of second-order
slow-roll expressions, as explained in the next paragraph.  In the
model $v=4$ the range for the lower derivatives of the potential is
slightly larger than in the models with $v<4$, which has its
repercussions embodied in slight degeneracies between the fourth
derivative and the lower derivatives.

Note that all figures containing information on the fourth derivative
of the potential contain only the model $v=4$ (red line) and not those
with $p=2$ or $p=3$, since in the slow-roll approximation one would
need to go to third order in order to infer $V''''_*$ from the
primordial spectrum.

Finally, it is worth pointing out that the results at all orders in
both parameterizations still allow for a flat (Harrison-Zel'dovich)
spectrum at the 95\%~C.L., or for a linear potential at the
68\%~C.L.

\vspace{0.5cm}

{\it Precision of the slow-roll approximation.}  In Fig.~\ref{fig6} we
show the discrepancy between the numerical results for the spectral
parameters (top left: $r$, top right: $n_S$, bottom left: $\alpha_S$)
and those obtained using the slow-roll approximation up to third order
in the derivatives of the inflaton potential (second order in
slow-roll parameters for $n_S$ and $\alpha_S$, third order for
$r$). The numerically calculated parameters can in this context be
treated as exact, since they do not involve any approximation (within
first-order cosmological perturbation theory), and remain perfectly
stable when we increase the precision parameters of our code. As $r$
naturally comes out at one order higher in the slow-roll expansion
than $n_S$ and $\alpha_S$, the top left plot ($r$) shows less
discrepancy than do the plots for $n_S$ and $\alpha_S$. However, for
large $r$ there is a clear deviation from slow-roll in the results
with $v=4$ (red), up to $\sim$7\%. In the plot for $n_S$ it is clearly
seen that second order slow roll is only accurate up to $\sim 3\%$ for
the run with $v=3$ (blue region). This discrepancy is important, since
the data constrains $n_S$ with a standard deviation $\sigma \sim
2$\%. When a fourth derivative of the inflaton potential is included,
second-order slow roll becomes really inaccurate, with a typical error
of 10\% on $n_S$, while the running can be wrongfully estimated by as
much as $\Delta \alpha_S=0.1$, i.e.  three standard deviations given
the current data.  In a future work, it would therefore be useful to
compute the next-order contributions to the running analytically (the
bottom right diagram shows the quasi-linear dependence of $\alpha_S$
on the combination $V''''_* {V'}_*^2/V_*^3$).

\section{Conclusions}

In this work, we derived some constraints on the inflaton potential
from up-to-date CMB and LSS data. Our CMB data consists in the WMAP
3-year measurement of the temperature and polarization power
spectrum. We did include the first (controversial) multipoles, after
checking in section~\ref{sec_spectrum} that they do not have a
significant impact on the determination of the primordial spectrum
tilt and running. Our analysis differs from previous works for several
reasons. First, we directly fit the parameters describing the inflaton
potential, instead of constraining first the primordial spectra, and
reconstructing the inflaton potential afterwards. Second, we
Taylor-expand the inflaton potential in the vicinity of the pivot
scale at a rather high order (up to $v=4$), and see that such a high
order is important e.g. for exploring all the parameter space allowed
by the data in terms of running of the scalar spectrum tilt. Third, we
compute the scalar and tensor spectra for each model numerically, and
find that for the models considered here this is important,
since the spectra derived from the second-order slow-roll formalism are
inaccurate by the same order as the observational constraints
themselves.

However, the most important peculiarity of this work is our choice to
focus only on the observable region of the inflaton potential, not
making any assumption on the shape of the potential between the
observable region and the minimum close to which inflaton stops after
approximately 50 e-folds (depending on the scale of inflation). This
choice has a crucial impact on the results. If we did extrapolate the
inflaton potential over 50 e-folds, keeping the same order in the
Taylor expansion, our models would be more severely constrained, since
the requirement of 50 extra e-folds would kill many of the allowed
potentials presented here\footnote{For instance, some limits on the
potential derivatives were presented up to $V'''$ in
\cite{Hansen:2001eu} and up to $V''''$ in \cite{Caprini:2002jy}. In
these works, most of the constraints on high derivatives come from the
requirement of at least 50 inflationary e-folds with the extrapolated
potential. Not surprisingly, the resulting bounds are much stronger
that ours.}.  We are perfectly aware of this, and wish to point out
that this is one of two points of view, which are both equally
sensible.

From one point of view, if one works under the prejudice that the
inflaton potential should not be too complicated, then it is extremely
relevant to consider the global shape of the potential and to throw
away all models which cannot sustain 60 inflationary e-folds. Many
papers use this approach, using sometimes Monte Carlo methods in which
the potential (or the Hubble flow $H(N)$) is Taylor expanded over the
60 e-folds at high order.

From another point of view, if one wants to address the question of what
is strictly allowed by the data, then even a high-order Taylor
expansion of the {\it full} potential sounds unsatisfactory for
modeling all its possible variations during such a long history as 60
e-folds (especially if one keeps in mind that some other fields could
then play a role: triggering a phase transition, inducing complicated
shapes as in the string-inspired landscape scenarios, etc.). On the
contrary, in this philosophy, one should only try to parametrize the
inflaton potential in the range probed by cosmological data,
i.e. around six or seven e-folds. This is what we did here, with a
Taylor expansion up to fourth order.

The two approaches lead, of course, to radically different
conclusions. For instance, in the first method, one would conclude
that during the observational e-folds the inflaton must be deep in the
slow-roll regime, since it is necessary to sustain a number of e-folds
which is an order of magnitude higher. The running would then be very
constrained~\cite{Easther:2006tv}.  In the second method, it is not a
problem to satisfy slow-roll only marginally on the edges of the
observable range. Even if $\epsilon_1$ grows dangerously close to one
when cluster scales exit the Hubble radius, the potential could become
much flatter afterwards, and sustain any desired amount of inflation.

Our main results for the inflaton potential reconstruction are
summarized in Figs.~\ref{fig4},~\ref{fig1},~\ref{fig2} and
Table~\ref{table2}. We also showed up to what extent the slow-roll
formalism reveals to be inaccurate in the current context in
Fig.~\ref{fig6}. This motivates possible future works concerning the
next-order slow-roll expressions.

Following the same approach, this work could be improved by adding
more large-scale structure data e.g. from Lyman-$\alpha$ forests or
weak lensing, which have a good power for further constraining the
primordial spectrum on smaller scales than the SDSS LRG data. Here we
choose to use a very restricted data set, in order
to derive rather conservative and robust results.

More generally, we point out that our {\sc cosmomc} module for
computing the primordial spectra numerically can be used in different
contexts, within {\sc cosmomc} or separately, and even (after minor
modifications) for studying more complicated models producing
characteristic features in the primordial spectra. The module was
written in a user-friendly way and made publicly available on the
website 
\url{http://wwwlapp.in2p3.fr/~lesgourgues/inflation/}.

\section*{Acknowledgements.}
This work was initiated during a very nice and fruitful stay at the
Galileo Galilei Institute for Theoretical Physics, supported by INFN.
The project was completed thanks to the support of the EU 6th
Framework Marie Curie Research and Training network ``UniverseNet''
(MRTN-CT-2006-035863). Numerical simulations were performed on the
PISTOO cluster of the IN2P3/CNRS Centre de Calcul (Lyon, France).

\bibliography{refs}

\begin{thebibliography}{70}
\expandafter\ifx\csname natexlab\endcsname\relax\def\natexlab#1{#1}\fi
\expandafter\ifx\csname bibnamefont\endcsname\relax
  \def\bibnamefont#1{#1}\fi
\expandafter\ifx\csname bibfnamefont\endcsname\relax
  \def\bibfnamefont#1{#1}\fi
\expandafter\ifx\csname citenamefont\endcsname\relax
  \def\citenamefont#1{#1}\fi
\expandafter\ifx\csname url\endcsname\relax
  \def\url#1{\texttt{#1}}\fi
\expandafter\ifx\csname urlprefix\endcsname\relax\def\urlprefix{URL }\fi
\providecommand{\bibinfo}[2]{#2}
\providecommand{\eprint}[2][]{\url{#2}}

\bibitem[{\citenamefont{Starobinsky}(1980)}]{Starobinsky:1980te}
\bibinfo{author}{\bibfnamefont{A.~A.} \bibnamefont{Starobinsky}},
  \bibinfo{journal}{Phys. Lett.} \textbf{\bibinfo{volume}{B91}},
  \bibinfo{pages}{99} (\bibinfo{year}{1980}).

\bibitem[{\citenamefont{Guth}(1981)}]{Guth:1980zm}
\bibinfo{author}{\bibfnamefont{A.~H.} \bibnamefont{Guth}},
  \bibinfo{journal}{Phys. Rev.} \textbf{\bibinfo{volume}{D23}},
  \bibinfo{pages}{347} (\bibinfo{year}{1981}).

\bibitem[{\citenamefont{Sato}(1981)}]{Sato:1980yn}
\bibinfo{author}{\bibfnamefont{K.}~\bibnamefont{Sato}}, \bibinfo{journal}{Mon.
  Not. Roy. Astron. Soc.} \textbf{\bibinfo{volume}{195}}, \bibinfo{pages}{467}
  (\bibinfo{year}{1981}).

\bibitem[{\citenamefont{Hawking and Moss}(1982)}]{Hawking:1981fz}
\bibinfo{author}{\bibfnamefont{S.~W.} \bibnamefont{Hawking}} \bibnamefont{and}
  \bibinfo{author}{\bibfnamefont{I.~G.} \bibnamefont{Moss}},
  \bibinfo{journal}{Phys. Lett.} \textbf{\bibinfo{volume}{B110}},
  \bibinfo{pages}{35} (\bibinfo{year}{1982}).

\bibitem[{\citenamefont{Linde}(1982{\natexlab{a}})}]{Linde:1981mu}
\bibinfo{author}{\bibfnamefont{A.~D.} \bibnamefont{Linde}},
  \bibinfo{journal}{Phys. Lett.} \textbf{\bibinfo{volume}{B108}},
  \bibinfo{pages}{389} (\bibinfo{year}{1982}{\natexlab{a}}).

\bibitem[{\citenamefont{Linde}(1983)}]{Linde:1983gd}
\bibinfo{author}{\bibfnamefont{A.~D.} \bibnamefont{Linde}},
  \bibinfo{journal}{Phys. Lett.} \textbf{\bibinfo{volume}{B129}},
  \bibinfo{pages}{177} (\bibinfo{year}{1983}).

\bibitem[{\citenamefont{Starobinsky}(1979)}]{Starobinsky:1979ty}
\bibinfo{author}{\bibfnamefont{A.~A.} \bibnamefont{Starobinsky}},
  \bibinfo{journal}{JETP Lett.} \textbf{\bibinfo{volume}{30}},
  \bibinfo{pages}{682} (\bibinfo{year}{1979}).

\bibitem[{\citenamefont{Hawking}(1982)}]{Hawking:1982cz}
\bibinfo{author}{\bibfnamefont{S.~W.} \bibnamefont{Hawking}},
  \bibinfo{journal}{Phys. Lett.} \textbf{\bibinfo{volume}{B115}},
  \bibinfo{pages}{295} (\bibinfo{year}{1982}).

\bibitem[{\citenamefont{Starobinsky}(1982)}]{Starobinsky:1982ee}
\bibinfo{author}{\bibfnamefont{A.~A.} \bibnamefont{Starobinsky}},
  \bibinfo{journal}{Phys. Lett.} \textbf{\bibinfo{volume}{B117}},
  \bibinfo{pages}{175} (\bibinfo{year}{1982}).

\bibitem[{\citenamefont{Guth and Pi}(1982)}]{Guth:1982ec}
\bibinfo{author}{\bibfnamefont{A.~H.} \bibnamefont{Guth}} \bibnamefont{and}
  \bibinfo{author}{\bibfnamefont{S.~Y.} \bibnamefont{Pi}},
  \bibinfo{journal}{Phys. Rev. Lett.} \textbf{\bibinfo{volume}{49}},
  \bibinfo{pages}{1110} (\bibinfo{year}{1982}).

\bibitem[{\citenamefont{Linde}(1982{\natexlab{b}})}]{Linde:1982uu}
\bibinfo{author}{\bibfnamefont{A.~D.} \bibnamefont{Linde}},
  \bibinfo{journal}{Phys. Lett.} \textbf{\bibinfo{volume}{B116}},
  \bibinfo{pages}{335} (\bibinfo{year}{1982}{\natexlab{b}}).

\bibitem[{\citenamefont{Bardeen et~al.}(1983)\citenamefont{Bardeen, Steinhardt,
  and Turner}}]{Bardeen:1983qw}
\bibinfo{author}{\bibfnamefont{J.~M.} \bibnamefont{Bardeen}},
  \bibinfo{author}{\bibfnamefont{P.~J.} \bibnamefont{Steinhardt}},
  \bibnamefont{and} \bibinfo{author}{\bibfnamefont{M.~S.}
  \bibnamefont{Turner}}, \bibinfo{journal}{Phys. Rev.}
  \textbf{\bibinfo{volume}{D28}}, \bibinfo{pages}{679} (\bibinfo{year}{1983}).

\bibitem[{\citenamefont{Abbott and Wise}(1984)}]{Abbott:1984fp}
\bibinfo{author}{\bibfnamefont{L.~F.} \bibnamefont{Abbott}} \bibnamefont{and}
  \bibinfo{author}{\bibfnamefont{M.~B.} \bibnamefont{Wise}},
  \bibinfo{journal}{Nucl. Phys.} \textbf{\bibinfo{volume}{B244}},
  \bibinfo{pages}{541} (\bibinfo{year}{1984}).

\bibitem[{\citenamefont{Salopek et~al.}(1989)\citenamefont{Salopek, Bond, and
  Bardeen}}]{Salopek:1988qh}
\bibinfo{author}{\bibfnamefont{D.~S.} \bibnamefont{Salopek}},
  \bibinfo{author}{\bibfnamefont{J.~R.} \bibnamefont{Bond}}, \bibnamefont{and}
  \bibinfo{author}{\bibfnamefont{J.~M.} \bibnamefont{Bardeen}},
  \bibinfo{journal}{Phys. Rev.} \textbf{\bibinfo{volume}{D40}},
  \bibinfo{pages}{1753} (\bibinfo{year}{1989}).

\bibitem[{\citenamefont{Spergel et~al.}(2006)}]{Spergel:2006hy}
\bibinfo{author}{\bibfnamefont{D.~N.} \bibnamefont{Spergel}}
  \bibnamefont{et~al.} (\bibinfo{year}{2006}), \eprint{astro-ph/0603449}.

\bibitem[{\citenamefont{Page et~al.}(2006)}]{Page:2006hz}
\bibinfo{author}{\bibfnamefont{L.}~\bibnamefont{Page}} \bibnamefont{et~al.}
  (\bibinfo{year}{2006}), \eprint{astro-ph/0603450}.

\bibitem[{\citenamefont{Hinshaw et~al.}(2006)}]{Hinshaw:2006ia}
\bibinfo{author}{\bibfnamefont{G.}~\bibnamefont{Hinshaw}} \bibnamefont{et~al.}
  (\bibinfo{year}{2006}), \eprint{astro-ph/0603451}.

\bibitem[{\citenamefont{Jarosik et~al.}(2006)}]{Jarosik:2006ib}
\bibinfo{author}{\bibfnamefont{N.}~\bibnamefont{Jarosik}} \bibnamefont{et~al.}
  (\bibinfo{year}{2006}), \eprint{astro-ph/0603452}.

\bibitem[{\citenamefont{Linde}(1991)}]{Linde:1991km}
\bibinfo{author}{\bibfnamefont{A.~D.} \bibnamefont{Linde}},
  \bibinfo{journal}{Phys. Lett.} \textbf{\bibinfo{volume}{B259}},
  \bibinfo{pages}{38} (\bibinfo{year}{1991}).

\bibitem[{\citenamefont{Linde}(1994)}]{Linde:1993cn}
\bibinfo{author}{\bibfnamefont{A.~D.} \bibnamefont{Linde}},
  \bibinfo{journal}{Phys. Rev.} \textbf{\bibinfo{volume}{D49}},
  \bibinfo{pages}{748} (\bibinfo{year}{1994}), \eprint{astro-ph/9307002}.

\bibitem[{\citenamefont{Copeland et~al.}(1994)\citenamefont{Copeland, Liddle,
  Lyth, Stewart, and Wands}}]{Copeland:1994vg}
\bibinfo{author}{\bibfnamefont{E.~J.} \bibnamefont{Copeland}},
  \bibinfo{author}{\bibfnamefont{A.~R.} \bibnamefont{Liddle}},
  \bibinfo{author}{\bibfnamefont{D.~H.} \bibnamefont{Lyth}},
  \bibinfo{author}{\bibfnamefont{E.~D.} \bibnamefont{Stewart}},
  \bibnamefont{and} \bibinfo{author}{\bibfnamefont{D.}~\bibnamefont{Wands}},
  \bibinfo{journal}{Phys. Rev.} \textbf{\bibinfo{volume}{D49}},
  \bibinfo{pages}{6410} (\bibinfo{year}{1994}), \eprint{astro-ph/9401011}.

\bibitem[{\citenamefont{Peiris and
  Easther}(2006{\natexlab{a}})}]{Peiris:2006ug}
\bibinfo{author}{\bibfnamefont{H.}~\bibnamefont{Peiris}} \bibnamefont{and}
  \bibinfo{author}{\bibfnamefont{R.}~\bibnamefont{Easther}},
  \bibinfo{journal}{JCAP} \textbf{\bibinfo{volume}{0607}}, \bibinfo{pages}{002}
  (\bibinfo{year}{2006}{\natexlab{a}}), \eprint{astro-ph/0603587}.

\bibitem[{\citenamefont{de~Vega and Sanchez}(2006)}]{deVega:2006hb}
\bibinfo{author}{\bibfnamefont{H.~J.} \bibnamefont{de~Vega}} \bibnamefont{and}
  \bibinfo{author}{\bibfnamefont{N.~G.} \bibnamefont{Sanchez}}
  (\bibinfo{year}{2006}), \eprint{astro-ph/0604136}.

\bibitem[{\citenamefont{Easther and Peiris}(2006)}]{Easther:2006tv}
\bibinfo{author}{\bibfnamefont{R.}~\bibnamefont{Easther}} \bibnamefont{and}
  \bibinfo{author}{\bibfnamefont{H.}~\bibnamefont{Peiris}},
  \bibinfo{journal}{JCAP} \textbf{\bibinfo{volume}{0609}}, \bibinfo{pages}{010}
  (\bibinfo{year}{2006}), \eprint{astro-ph/0604214}.

\bibitem[{\citenamefont{Kinney et~al.}(2006)\citenamefont{Kinney, Kolb,
  Melchiorri, and Riotto}}]{Kinney:2006qm}
\bibinfo{author}{\bibfnamefont{W.~H.} \bibnamefont{Kinney}},
  \bibinfo{author}{\bibfnamefont{E.~W.} \bibnamefont{Kolb}},
  \bibinfo{author}{\bibfnamefont{A.}~\bibnamefont{Melchiorri}},
  \bibnamefont{and} \bibinfo{author}{\bibfnamefont{A.}~\bibnamefont{Riotto}},
  \bibinfo{journal}{Phys. Rev.} \textbf{\bibinfo{volume}{D74}},
  \bibinfo{pages}{023502} (\bibinfo{year}{2006}), \eprint{astro-ph/0605338}.

\bibitem[{\citenamefont{Martin and Ringeval}(2006)}]{Martin:2006rs}
\bibinfo{author}{\bibfnamefont{J.}~\bibnamefont{Martin}} \bibnamefont{and}
  \bibinfo{author}{\bibfnamefont{C.}~\bibnamefont{Ringeval}},
  \bibinfo{journal}{JCAP} \textbf{\bibinfo{volume}{0608}}, \bibinfo{pages}{009}
  (\bibinfo{year}{2006}), \eprint{astro-ph/0605367}.

\bibitem[{\citenamefont{Covi et~al.}(2006)\citenamefont{Covi, Hamann,
  Melchiorri, Slosar, and Sorbera}}]{Covi:2006ci}
\bibinfo{author}{\bibfnamefont{L.}~\bibnamefont{Covi}},
  \bibinfo{author}{\bibfnamefont{J.}~\bibnamefont{Hamann}},
  \bibinfo{author}{\bibfnamefont{A.}~\bibnamefont{Melchiorri}},
  \bibinfo{author}{\bibfnamefont{A.}~\bibnamefont{Slosar}}, \bibnamefont{and}
  \bibinfo{author}{\bibfnamefont{I.}~\bibnamefont{Sorbera}},
  \bibinfo{journal}{Phys. Rev.} \textbf{\bibinfo{volume}{D74}},
  \bibinfo{pages}{083509} (\bibinfo{year}{2006}), \eprint{astro-ph/0606452}.

\bibitem[{\citenamefont{Finelli et~al.}(2006)\citenamefont{Finelli, Rianna, and
  Mandolesi}}]{Finelli:2006fi}
\bibinfo{author}{\bibfnamefont{F.}~\bibnamefont{Finelli}},
  \bibinfo{author}{\bibfnamefont{M.}~\bibnamefont{Rianna}}, \bibnamefont{and}
  \bibinfo{author}{\bibfnamefont{N.}~\bibnamefont{Mandolesi}},
  \bibinfo{journal}{JCAP} \textbf{\bibinfo{volume}{0612}}, \bibinfo{pages}{006}
  (\bibinfo{year}{2006}), \eprint{astro-ph/0608277}.

\bibitem[{\citenamefont{Peiris and
  Easther}(2006{\natexlab{b}})}]{Peiris:2006sj}
\bibinfo{author}{\bibfnamefont{H.}~\bibnamefont{Peiris}} \bibnamefont{and}
  \bibinfo{author}{\bibfnamefont{R.}~\bibnamefont{Easther}},
  \bibinfo{journal}{JCAP} \textbf{\bibinfo{volume}{0610}}, \bibinfo{pages}{017}
  (\bibinfo{year}{2006}{\natexlab{b}}), \eprint{astro-ph/0609003}.

\bibitem[{\citenamefont{Destri et~al.}(2007)\citenamefont{Destri, de~Vega, and
  Sanchez}}]{Destri:2007pv}
\bibinfo{author}{\bibfnamefont{C.}~\bibnamefont{Destri}},
  \bibinfo{author}{\bibfnamefont{H.~J.} \bibnamefont{de~Vega}},
  \bibnamefont{and} \bibinfo{author}{\bibfnamefont{N.~G.}
  \bibnamefont{Sanchez}} (\bibinfo{year}{2007}), \eprint{astro-ph/0703417}.

\bibitem[{\citenamefont{Ringeval}(2007)}]{Ringeval:2007am}
\bibinfo{author}{\bibfnamefont{C.}~\bibnamefont{Ringeval}}
  (\bibinfo{year}{2007}), \eprint{astro-ph/0703486}.

\bibitem[{\citenamefont{Cardoso}(2007)}]{Cardoso:2006wf}
\bibinfo{author}{\bibfnamefont{A.}~\bibnamefont{Cardoso}},
  \bibinfo{journal}{Phys. Rev.} \textbf{\bibinfo{volume}{D75}},
  \bibinfo{pages}{027302} (\bibinfo{year}{2007}), \eprint{astro-ph/0610074}.

\bibitem[{\citenamefont{Cline and Hoi}(2006)}]{Cline:2006db}
\bibinfo{author}{\bibfnamefont{J.~M.} \bibnamefont{Cline}} \bibnamefont{and}
  \bibinfo{author}{\bibfnamefont{L.}~\bibnamefont{Hoi}},
  \bibinfo{journal}{JCAP} \textbf{\bibinfo{volume}{0606}}, \bibinfo{pages}{007}
  (\bibinfo{year}{2006}), \eprint{astro-ph/0603403}.

\bibitem[{\citenamefont{Grivell and Liddle}(2000)}]{Grivell:1999wc}
\bibinfo{author}{\bibfnamefont{I.~J.} \bibnamefont{Grivell}} \bibnamefont{and}
  \bibinfo{author}{\bibfnamefont{A.~R.} \bibnamefont{Liddle}},
  \bibinfo{journal}{Phys. Rev.} \textbf{\bibinfo{volume}{D61}},
  \bibinfo{pages}{081301} (\bibinfo{year}{2000}), \eprint{astro-ph/9906327}.

\bibitem[{\citenamefont{Steinhardt and Turner}(1984)}]{Steinhardt:1984jj}
\bibinfo{author}{\bibfnamefont{P.~J.} \bibnamefont{Steinhardt}}
  \bibnamefont{and} \bibinfo{author}{\bibfnamefont{M.~S.}
  \bibnamefont{Turner}}, \bibinfo{journal}{Phys. Rev.}
  \textbf{\bibinfo{volume}{D29}}, \bibinfo{pages}{2162} (\bibinfo{year}{1984}).

\bibitem[{\citenamefont{Salopek and Bond}(1990)}]{Salopek:1990jq}
\bibinfo{author}{\bibfnamefont{D.~S.} \bibnamefont{Salopek}} \bibnamefont{and}
  \bibinfo{author}{\bibfnamefont{J.~R.} \bibnamefont{Bond}},
  \bibinfo{journal}{Phys. Rev.} \textbf{\bibinfo{volume}{D42}},
  \bibinfo{pages}{3936} (\bibinfo{year}{1990}).

\bibitem[{\citenamefont{Liddle et~al.}(1994)\citenamefont{Liddle, Parsons, and
  Barrow}}]{Liddle:1994dx}
\bibinfo{author}{\bibfnamefont{A.~R.} \bibnamefont{Liddle}},
  \bibinfo{author}{\bibfnamefont{P.}~\bibnamefont{Parsons}}, \bibnamefont{and}
  \bibinfo{author}{\bibfnamefont{J.~D.} \bibnamefont{Barrow}},
  \bibinfo{journal}{Phys. Rev.} \textbf{\bibinfo{volume}{D50}},
  \bibinfo{pages}{7222} (\bibinfo{year}{1994}), \eprint{astro-ph/9408015}.

\bibitem[{\citenamefont{Stewart and Lyth}(1993)}]{Stewart:1993bc}
\bibinfo{author}{\bibfnamefont{E.~D.} \bibnamefont{Stewart}} \bibnamefont{and}
  \bibinfo{author}{\bibfnamefont{D.~H.} \bibnamefont{Lyth}},
  \bibinfo{journal}{Phys. Lett.} \textbf{\bibinfo{volume}{B302}},
  \bibinfo{pages}{171} (\bibinfo{year}{1993}), \eprint{gr-qc/9302019}.

\bibitem[{\citenamefont{Lidsey et~al.}(1997)}]{Lidsey:1995np}
\bibinfo{author}{\bibfnamefont{J.~E.} \bibnamefont{Lidsey}}
  \bibnamefont{et~al.}, \bibinfo{journal}{Rev. Mod. Phys.}
  \textbf{\bibinfo{volume}{69}}, \bibinfo{pages}{373} (\bibinfo{year}{1997}),
  \eprint{astro-ph/9508078}.

\bibitem[{\citenamefont{Gong and Stewart}(2001)}]{Gong:2001he}
\bibinfo{author}{\bibfnamefont{J.-O.} \bibnamefont{Gong}} \bibnamefont{and}
  \bibinfo{author}{\bibfnamefont{E.~D.} \bibnamefont{Stewart}},
  \bibinfo{journal}{Phys. Lett.} \textbf{\bibinfo{volume}{B510}},
  \bibinfo{pages}{1} (\bibinfo{year}{2001}), \eprint{astro-ph/0101225}.

\bibitem[{\citenamefont{Dodelson and Stewart}(2002)}]{Dodelson:2001sh}
\bibinfo{author}{\bibfnamefont{S.}~\bibnamefont{Dodelson}} \bibnamefont{and}
  \bibinfo{author}{\bibfnamefont{E.}~\bibnamefont{Stewart}},
  \bibinfo{journal}{Phys. Rev.} \textbf{\bibinfo{volume}{D65}},
  \bibinfo{pages}{101301} (\bibinfo{year}{2002}), \eprint{astro-ph/0109354}.

\bibitem[{\citenamefont{Schwarz et~al.}(2001)\citenamefont{Schwarz,
  Terrero-Escalante, and Garcia}}]{Schwarz:2001vv}
\bibinfo{author}{\bibfnamefont{D.~J.} \bibnamefont{Schwarz}},
  \bibinfo{author}{\bibfnamefont{C.~A.} \bibnamefont{Terrero-Escalante}},
  \bibnamefont{and} \bibinfo{author}{\bibfnamefont{A.~A.}
  \bibnamefont{Garcia}}, \bibinfo{journal}{Phys. Lett.}
  \textbf{\bibinfo{volume}{B517}}, \bibinfo{pages}{243} (\bibinfo{year}{2001}),
  \eprint{astro-ph/0106020}.

\bibitem[{\citenamefont{Stewart}(2002)}]{Stewart:2001cd}
\bibinfo{author}{\bibfnamefont{E.~D.} \bibnamefont{Stewart}},
  \bibinfo{journal}{Phys. Rev.} \textbf{\bibinfo{volume}{D65}},
  \bibinfo{pages}{103508} (\bibinfo{year}{2002}), \eprint{astro-ph/0110322}.

\bibitem[{\citenamefont{Leach et~al.}(2002)\citenamefont{Leach, Liddle, Martin,
  and Schwarz}}]{Leach:2002ar}
\bibinfo{author}{\bibfnamefont{S.~M.} \bibnamefont{Leach}},
  \bibinfo{author}{\bibfnamefont{A.~R.} \bibnamefont{Liddle}},
  \bibinfo{author}{\bibfnamefont{J.}~\bibnamefont{Martin}}, \bibnamefont{and}
  \bibinfo{author}{\bibfnamefont{D.~J.} \bibnamefont{Schwarz}},
  \bibinfo{journal}{Phys. Rev.} \textbf{\bibinfo{volume}{D66}},
  \bibinfo{pages}{023515} (\bibinfo{year}{2002}), \eprint{astro-ph/0202094}.

\bibitem[{\citenamefont{Hansen and Kunz}(2002)}]{Hansen:2001eu}
\bibinfo{author}{\bibfnamefont{S.~H.} \bibnamefont{Hansen}} \bibnamefont{and}
  \bibinfo{author}{\bibfnamefont{M.}~\bibnamefont{Kunz}},
  \bibinfo{journal}{Mon. Not. Roy. Astron. Soc.}
  \textbf{\bibinfo{volume}{336}}, \bibinfo{pages}{1007} (\bibinfo{year}{2002}),
  \eprint{hep-ph/0109252}.

\bibitem[{\citenamefont{Caprini et~al.}(2003)\citenamefont{Caprini, Hansen, and
  Kunz}}]{Caprini:2002jy}
\bibinfo{author}{\bibfnamefont{C.}~\bibnamefont{Caprini}},
  \bibinfo{author}{\bibfnamefont{S.~H.} \bibnamefont{Hansen}},
  \bibnamefont{and} \bibinfo{author}{\bibfnamefont{M.}~\bibnamefont{Kunz}},
  \bibinfo{journal}{Mon. Not. Roy. Astron. Soc.}
  \textbf{\bibinfo{volume}{339}}, \bibinfo{pages}{212} (\bibinfo{year}{2003}),
  \eprint{hep-ph/0210095}.

\bibitem[{\citenamefont{Liddle}(2003)}]{Liddle:2003py}
\bibinfo{author}{\bibfnamefont{A.~R.} \bibnamefont{Liddle}},
  \bibinfo{journal}{Phys. Rev.} \textbf{\bibinfo{volume}{D68}},
  \bibinfo{pages}{103504} (\bibinfo{year}{2003}), \eprint{astro-ph/0307286}.

\bibitem[{\citenamefont{Choe et~al.}(2004)\citenamefont{Choe, Gong, and
  Stewart}}]{Choe:2004zg}
\bibinfo{author}{\bibfnamefont{J.}~\bibnamefont{Choe}},
  \bibinfo{author}{\bibfnamefont{J.-O.} \bibnamefont{Gong}}, \bibnamefont{and}
  \bibinfo{author}{\bibfnamefont{E.~D.} \bibnamefont{Stewart}},
  \bibinfo{journal}{JCAP} \textbf{\bibinfo{volume}{0407}}, \bibinfo{pages}{012}
  (\bibinfo{year}{2004}), \eprint{hep-ph/0405155}.

\bibitem[{\citenamefont{Habib et~al.}(2005)\citenamefont{Habib, Heinen,
  Heitmann, and Jungman}}]{Habib:2005mh}
\bibinfo{author}{\bibfnamefont{S.}~\bibnamefont{Habib}},
  \bibinfo{author}{\bibfnamefont{A.}~\bibnamefont{Heinen}},
  \bibinfo{author}{\bibfnamefont{K.}~\bibnamefont{Heitmann}}, \bibnamefont{and}
  \bibinfo{author}{\bibfnamefont{G.}~\bibnamefont{Jungman}},
  \bibinfo{journal}{Phys. Rev.} \textbf{\bibinfo{volume}{D71}},
  \bibinfo{pages}{043518} (\bibinfo{year}{2005}), \eprint{astro-ph/0501130}.

\bibitem[{\citenamefont{Joy et~al.}(2005)\citenamefont{Joy, Stewart, Gong, and
  Lee}}]{Joy:2005ep}
\bibinfo{author}{\bibfnamefont{M.}~\bibnamefont{Joy}},
  \bibinfo{author}{\bibfnamefont{E.~D.} \bibnamefont{Stewart}},
  \bibinfo{author}{\bibfnamefont{J.-O.} \bibnamefont{Gong}}, \bibnamefont{and}
  \bibinfo{author}{\bibfnamefont{H.-C.} \bibnamefont{Lee}},
  \bibinfo{journal}{JCAP} \textbf{\bibinfo{volume}{0504}}, \bibinfo{pages}{012}
  (\bibinfo{year}{2005}), \eprint{astro-ph/0501659}.

\bibitem[{\citenamefont{Kadota et~al.}(2005)\citenamefont{Kadota, Dodelson, Hu,
  and Stewart}}]{Kadota:2005hv}
\bibinfo{author}{\bibfnamefont{K.}~\bibnamefont{Kadota}},
  \bibinfo{author}{\bibfnamefont{S.}~\bibnamefont{Dodelson}},
  \bibinfo{author}{\bibfnamefont{W.}~\bibnamefont{Hu}}, \bibnamefont{and}
  \bibinfo{author}{\bibfnamefont{E.~D.} \bibnamefont{Stewart}},
  \bibinfo{journal}{Phys. Rev.} \textbf{\bibinfo{volume}{D72}},
  \bibinfo{pages}{023510} (\bibinfo{year}{2005}), \eprint{astro-ph/0505158}.

\bibitem[{\citenamefont{Casadio et~al.}(2006)\citenamefont{Casadio, Finelli,
  Kamenshchik, Luzzi, and Venturi}}]{Casadio:2006wb}
\bibinfo{author}{\bibfnamefont{R.}~\bibnamefont{Casadio}},
  \bibinfo{author}{\bibfnamefont{F.}~\bibnamefont{Finelli}},
  \bibinfo{author}{\bibfnamefont{A.}~\bibnamefont{Kamenshchik}},
  \bibinfo{author}{\bibfnamefont{M.}~\bibnamefont{Luzzi}}, \bibnamefont{and}
  \bibinfo{author}{\bibfnamefont{G.}~\bibnamefont{Venturi}},
  \bibinfo{journal}{JCAP} \textbf{\bibinfo{volume}{0604}}, \bibinfo{pages}{011}
  (\bibinfo{year}{2006}), \eprint{gr-qc/0603026}.

\bibitem[{\citenamefont{de~Oliveira and
  Terrero-Escalante}(2006)}]{deOliveira:2005mf}
\bibinfo{author}{\bibfnamefont{H.~P.} \bibnamefont{de~Oliveira}}
  \bibnamefont{and} \bibinfo{author}{\bibfnamefont{C.~A.}
  \bibnamefont{Terrero-Escalante}}, \bibinfo{journal}{JCAP}
  \textbf{\bibinfo{volume}{0601}}, \bibinfo{pages}{024} (\bibinfo{year}{2006}),
  \eprint{astro-ph/0511660}.

\bibitem[{\citenamefont{Makarov}(2005)}]{Makarov:2005uh}
\bibinfo{author}{\bibfnamefont{A.}~\bibnamefont{Makarov}},
  \bibinfo{journal}{Phys. Rev.} \textbf{\bibinfo{volume}{D72}},
  \bibinfo{pages}{083517} (\bibinfo{year}{2005}), \eprint{astro-ph/0506326}.

\bibitem[{\citenamefont{Tegmark et~al.}(2006)}]{Tegmark:2006az}
\bibinfo{author}{\bibfnamefont{M.}~\bibnamefont{Tegmark}} \bibnamefont{et~al.},
  \bibinfo{journal}{Phys. Rev.} \textbf{\bibinfo{volume}{D74}},
  \bibinfo{pages}{123507} (\bibinfo{year}{2006}), \eprint{astro-ph/0608632}.

\bibitem[{\citenamefont{Beltran et~al.}(2005)\citenamefont{Beltran,
  Garcia-Bellido, Lesgourgues, Liddle, and Slosar}}]{Beltran:2005xd}
\bibinfo{author}{\bibfnamefont{M.}~\bibnamefont{Beltran}},
  \bibinfo{author}{\bibfnamefont{J.}~\bibnamefont{Garcia-Bellido}},
  \bibinfo{author}{\bibfnamefont{J.}~\bibnamefont{Lesgourgues}},
  \bibinfo{author}{\bibfnamefont{A.~R.} \bibnamefont{Liddle}},
  \bibnamefont{and} \bibinfo{author}{\bibfnamefont{A.}~\bibnamefont{Slosar}},
  \bibinfo{journal}{Phys. Rev.} \textbf{\bibinfo{volume}{D71}},
  \bibinfo{pages}{063532} (\bibinfo{year}{2005}), \eprint{astro-ph/0501477}.

\bibitem[{\citenamefont{Trotta}(2005)}]{Trotta:2005ar}
\bibinfo{author}{\bibfnamefont{R.}~\bibnamefont{Trotta}}
  (\bibinfo{year}{2005}), \eprint{astro-ph/0504022}.

\bibitem[{\citenamefont{Kunz et~al.}(2006)\citenamefont{Kunz, Trotta, and
  Parkinson}}]{Kunz:2006mc}
\bibinfo{author}{\bibfnamefont{M.}~\bibnamefont{Kunz}},
  \bibinfo{author}{\bibfnamefont{R.}~\bibnamefont{Trotta}}, \bibnamefont{and}
  \bibinfo{author}{\bibfnamefont{D.}~\bibnamefont{Parkinson}},
  \bibinfo{journal}{Phys. Rev.} \textbf{\bibinfo{volume}{D74}},
  \bibinfo{pages}{023503} (\bibinfo{year}{2006}), \eprint{astro-ph/0602378}.

\bibitem[{\citenamefont{Parkinson et~al.}(2006)\citenamefont{Parkinson,
  Mukherjee, and Liddle}}]{Parkinson:2006ku}
\bibinfo{author}{\bibfnamefont{D.}~\bibnamefont{Parkinson}},
  \bibinfo{author}{\bibfnamefont{P.}~\bibnamefont{Mukherjee}},
  \bibnamefont{and} \bibinfo{author}{\bibfnamefont{A.~R.}
  \bibnamefont{Liddle}}, \bibinfo{journal}{Phys. Rev.}
  \textbf{\bibinfo{volume}{D73}}, \bibinfo{pages}{123523}
  (\bibinfo{year}{2006}), \eprint{astro-ph/0605003}.

\bibitem[{\citenamefont{Pahud et~al.}(2006)\citenamefont{Pahud, Liddle,
  Mukherjee, and Parkinson}}]{Pahud:2006kv}
\bibinfo{author}{\bibfnamefont{C.}~\bibnamefont{Pahud}},
  \bibinfo{author}{\bibfnamefont{A.~R.} \bibnamefont{Liddle}},
  \bibinfo{author}{\bibfnamefont{P.}~\bibnamefont{Mukherjee}},
  \bibnamefont{and}
  \bibinfo{author}{\bibfnamefont{D.}~\bibnamefont{Parkinson}},
  \bibinfo{journal}{Phys. Rev.} \textbf{\bibinfo{volume}{D73}},
  \bibinfo{pages}{123524} (\bibinfo{year}{2006}), \eprint{astro-ph/0605004}.

\bibitem[{\citenamefont{Liddle et~al.}(2006)\citenamefont{Liddle, Mukherjee,
  and Parkinson}}]{Liddle:2006tc}
\bibinfo{author}{\bibfnamefont{A.~R.} \bibnamefont{Liddle}},
  \bibinfo{author}{\bibfnamefont{P.}~\bibnamefont{Mukherjee}},
  \bibnamefont{and}
  \bibinfo{author}{\bibfnamefont{D.}~\bibnamefont{Parkinson}},
  \bibinfo{journal}{Astron. Geophys.} \textbf{\bibinfo{volume}{47}},
  \bibinfo{pages}{4.30} (\bibinfo{year}{2006}), \eprint{astro-ph/0608184}.

\bibitem[{\citenamefont{Liddle}(2007)}]{Liddle:2007fy}
\bibinfo{author}{\bibfnamefont{A.~R.} \bibnamefont{Liddle}}
  (\bibinfo{year}{2007}), \eprint{astro-ph/0701113}.

\bibitem[{\citenamefont{Leach}(2006)}]{Leach:2005av}
\bibinfo{author}{\bibfnamefont{S.}~\bibnamefont{Leach}}, \bibinfo{journal}{Mon.
  Not. Roy. Astron. Soc.} \textbf{\bibinfo{volume}{372}}, \bibinfo{pages}{646}
  (\bibinfo{year}{2006}), \eprint{astro-ph/0506390}.

\bibitem[{\citenamefont{Lewis and Bridle}(2002)}]{Lewis:2002ah}
\bibinfo{author}{\bibfnamefont{A.}~\bibnamefont{Lewis}} \bibnamefont{and}
  \bibinfo{author}{\bibfnamefont{S.}~\bibnamefont{Bridle}},
  \bibinfo{journal}{Phys. Rev.} \textbf{\bibinfo{volume}{D66}},
  \bibinfo{pages}{103511} (\bibinfo{year}{2002}), \eprint{astro-ph/0205436}.

\bibitem[{\citenamefont{Schwarz et~al.}(2004)\citenamefont{Schwarz, Starkman,
  Huterer, and Copi}}]{Schwarz:2004gk}
\bibinfo{author}{\bibfnamefont{D.~J.} \bibnamefont{Schwarz}},
  \bibinfo{author}{\bibfnamefont{G.~D.} \bibnamefont{Starkman}},
  \bibinfo{author}{\bibfnamefont{D.}~\bibnamefont{Huterer}}, \bibnamefont{and}
  \bibinfo{author}{\bibfnamefont{C.~J.} \bibnamefont{Copi}},
  \bibinfo{journal}{Phys. Rev. Lett.} \textbf{\bibinfo{volume}{93}},
  \bibinfo{pages}{221301} (\bibinfo{year}{2004}), \eprint{astro-ph/0403353}.

\bibitem[{\citenamefont{Copi et~al.}(2006)\citenamefont{Copi, Huterer, Schwarz,
  and Starkman}}]{Copi:2005ff}
\bibinfo{author}{\bibfnamefont{C.~J.} \bibnamefont{Copi}},
  \bibinfo{author}{\bibfnamefont{D.}~\bibnamefont{Huterer}},
  \bibinfo{author}{\bibfnamefont{D.~J.} \bibnamefont{Schwarz}},
  \bibnamefont{and} \bibinfo{author}{\bibfnamefont{G.~D.}
  \bibnamefont{Starkman}}, \bibinfo{journal}{Mon. Not. Roy. Astron. Soc.}
  \textbf{\bibinfo{volume}{367}}, \bibinfo{pages}{79} (\bibinfo{year}{2006}),
  \eprint{astro-ph/0508047}.

\bibitem[{\citenamefont{Copi et~al.}(2007)\citenamefont{Copi, Huterer, Schwarz,
  and Starkman}}]{Copi:2006tu}
\bibinfo{author}{\bibfnamefont{C.}~\bibnamefont{Copi}},
  \bibinfo{author}{\bibfnamefont{D.}~\bibnamefont{Huterer}},
  \bibinfo{author}{\bibfnamefont{D.}~\bibnamefont{Schwarz}}, \bibnamefont{and}
  \bibinfo{author}{\bibfnamefont{G.}~\bibnamefont{Starkman}},
  \bibinfo{journal}{Phys. Rev.} \textbf{\bibinfo{volume}{D75}},
  \bibinfo{pages}{023507} (\bibinfo{year}{2007}), \eprint{astro-ph/0605135}.

\bibitem[{\citenamefont{Jones et~al.}(2006)}]{Jones:2005yb}
\bibinfo{author}{\bibfnamefont{W.~C.} \bibnamefont{Jones}}
  \bibnamefont{et~al.}, \bibinfo{journal}{Astrophys. J.}
  \textbf{\bibinfo{volume}{647}}, \bibinfo{pages}{823} (\bibinfo{year}{2006}),
  \eprint{astro-ph/0507494}.

\bibitem[{\citenamefont{Kuo et~al.}(2004)}]{Kuo:2002ua}
\bibinfo{author}{\bibfnamefont{C.-l.} \bibnamefont{Kuo}} \bibnamefont{et~al.}
  (\bibinfo{collaboration}{ACBAR}), \bibinfo{journal}{Astrophys. J.}
  \textbf{\bibinfo{volume}{600}}, \bibinfo{pages}{32} (\bibinfo{year}{2004}),
  \eprint{astro-ph/0212289}.

\bibitem[{\citenamefont{Sievers et~al.}(2005)}]{Sievers:2005gj}
\bibinfo{author}{\bibfnamefont{J.~L.} \bibnamefont{Sievers}}
  \bibnamefont{et~al.} (\bibinfo{year}{2005}), \eprint{astro-ph/0509203}.

\end{thebibliography}

\end{document}